\documentclass[aps,prb,twocolumn,groupedaddress,showpacs]{revtex4}
\def\bg#1{\mbox{\boldmath $#1$}}

\usepackage{amsmath}
\usepackage{amssymb}
\usepackage{graphicx}
\usepackage{dsfont}
\usepackage{color}

\begin{document}

\title{Anisotropic magnetoresistance of spin-orbit coupled
carriers scattered from polarized magnetic impurities}

\author{Maxim Trushin$^{1,2}$, Karel V\'yborn\'y$^3$,
Peter Moraczewski$^4$,  Alexey A. Kovalev$^5$,
John Schliemann$^1$, and T. Jungwirth$^{3,6}$}

\address{$^1$Institut f\"ur Theoretische Physik, Universit\"at Regensburg,
D-93040 Regensburg, Germany}
\address{$^2$Physics Department, University of Texas, 1 University Station
  C1600, Austin, 78712 Texas, USA}
\address{$^3$Institute of Physics,  Academy of Sciences of the 
Czech Rep., v.v.i., Cukrovarnick\'a 10, Praha 6 CZ-16253, Czech Republic}
\address{$^4$I. Institut f\"ur Theoretische Physik, Universit\"at Hamburg, 
Jungiusstrasse 9, 20355 Hamburg, Germany}
\address{$^5$Department of Physics and Astronomy, University of California,
Los Angeles, 90095 California, USA}
\address{$^6$School of Physics and Astronomy, University of Nottingham, 
Nottingham NG7 2RD, United Kingdom}

\date{Sep04, 2009}

\begin{abstract}
  Anisotropic magnetoresistance (AMR) is a relativistic magnetotransport
  phenomenon arising from combined effects of spin-orbit coupling and broken
  symmetry of a ferromagnetically ordered state of the system.  In this work
  we focus on one realization of the AMR in which spin-orbit coupling enters
  via specific spin-textures on the carrier Fermi surfaces and ferromagnetism
  via elastic 
  scattering of carriers from polarized magnetic impurities. We report
  detailed heuristic examination, using model spin-orbit coupled systems, of
  the emergence of positive AMR (maximum resistivity for magnetization along
  current), negative AMR (minimum resistivity for magnetization along
  current), and of the crystalline AMR (resistivity depends on the absolute
  orientation of the magnetization and current vectors with respect to the
  crystal axes) components. We emphasize potential qualitative differences
  between pure magnetic and combined electro-magnetic impurity potentials,
  between short-range and long-range impurities, and between spin-1/2 and
  higher spin-state carriers.
  Conclusions based on our heuristic analysis are supported by exact
  solutions to the integral form of
  the Boltzmann transport equation in archetypical two-dimensional
  electron systems with Rashba
  and Dresselhaus spin-orbit interactions and in the three-dimensional
  spherical Kohn-Littinger model.
  We include comments on the relation of
  our microscopic calculations to standard phenomenology of the full angular
  dependence of the AMR, and on the relevance of our study to realistic,
  two-dimensional conduction-band carrier systems and to anisotropic transport
  in the valence band of diluted magnetic semiconductors.
\end{abstract}


%%% PACS
% 72.10.-d      Theory of electronic transport; scattering mechanisms
% 72.20.My      Galvanomagnetic and other magnetotransport effects
% 72.25.Dc      Spin polarized transport in semiconductors

\maketitle

\section{Introduction}
Advanced theoretical approaches and experiments in new unconventional
ferromagnets have recently led to a renewed interest in the relativistic,
extraordinary magnetotransport effects. There are two distinct extraordinary
magnetoresistance coefficients, the anomalous Hall
effect (AHE) and the anisotropic
magnetoresistance (AMR).
The AHE is the antisymmetric transverse magnetoresistance coefficient obeying
$\rho_{xy}(\mathbf{M})=-\rho_{xy}(-\mathbf{M})$,
where the magnetization vector $\mathbf{M}$ is pointing perpendicular to
the $\hat{x},\hat{y}$ plane of a Hall bar sample.  The AMR is the symmetric
coefficient with the longitudinal and transverse resistivities obeying,
$\rho_{xx}(\mathbf{M})=\rho_{xx}(-\mathbf{M})$ and
$\rho_{xy}(\mathbf{M})=\rho_{xy}(-\mathbf{M})$, where $\mathbf{M}$ has
an arbitrary orientation but in most studies it lies in the $x-y$
plane.  Numerous works have explored the origins of the  AHE; for reviews see
e.g. Refs.~\onlinecite{Chien:1980_a,Sinova:2004_c,Sinitsyn:2007_a}. Diluted
magnetic semiconductors became one of the favorable test bed systems for AHE
investigation\cite{Jungwirth:2002_a,Dietl:2003_c,Ruzmetov:2004_a,Chun:2006_a,Mihaly:2007_a,Arsenault:2008_b,Pu:2008_a} 
due to their tunability and the relatively simple, yet strongly spin-orbit
coupled Fermi surfaces.\cite{Matsukura:2002_a,Jungwirth:2006_a} An even more
systematic and comprehensive understanding of the AHE on a model level has
been obtained by considering two-dimensional semiconductor systems
with archetypical spin-orbit interactions (SOIs) of the Rashba and Dresselhaus
type.\cite{Dugaev:2005_a,Trushin:2006_a,Sinitsyn:2006_b,Inoue:2006_a,Onoda:2007_a,Nunner:2007_a,Nunner:2007_b,Borunda:2007_a,Sinitsyn:2007_a,Kovalev:2008_a,Kovalev:2009_a,Sinova:2008_a}

Despite the long history and importance in magnetic recording technologies,
the AMR has been studied less extensively.\cite{Smit:1951_a,Jaoul:1977_a,McGuire:1975_a,Banhart:1995_a,Khmelevskyi:2003_a}
Similar to the AHE, it has been recently argued that the analysis of the AMR
can be significantly simplified in diluted magnetic semiconductors like
(Ga,Mn)As.\cite{Rushforth:2007_a,Jungwirth:2008_a} Two distinct microscopic
mechanisms have been identified that can lead to anisotropic carrier life-times
in these systems: One combines the spin-orbit coupling in an unpolarized
carrier band with scattering off polarized magnetic impurities while the other
emphasizes polarization of the carrier band itself and does not require
magnetic nature of the scatterers. (Note that apart from life-times, 
the AMR may also arise from
anisotropic group velocities.\footnote{This combined effect of SOI and carrier
  polarization is the third type of microscopic mechanism that can lead to
  AMR. In (Ga,Mn)As, however, carrier-polarization-related anisotropy in group
  velocities is also weak, see
  Refs.~\onlinecite{Rushforth:2007_a,Rushforth:2007_b,Vyborny:2009_a}.})   
Although acting simultaneously in real systems, theoretically both mechanisms
can be turned on and off independently and it was found\cite{Rushforth:2007_a}
that the scattering of spin-orbit coupled band carriers from magnetically
polarized impurities should dominate in the diluted magnetic
semiconductors. Building on the analogy with AHE studies we seek further
insight into the basic physics of this AMR mechanism by focusing on the
archetypical spin-orbit coupled two-dimensional systems. 

Using the relaxation-time approximation (RTA) and starting with the Rashba and
Dresselhaus models we show in Sec.~\ref{sec-II} how the sign of the AMR can be
inferred by inspecting the spin texture of the spin-orbit coupled Fermi
surface. We point out that impurities containing polarized magnetic potential
only or containing a combined electro-magnetic potential can yield distinct
AMR phenomenologies. Examination of the Rashba and Dresselhaus models allows
us to draw separate links between the spin-texture and the non-crystalline and
crystalline AMR components where the non-crystalline AMR depends on the
relative angle between ${\bf M}$ and current ${\bf I}$ while the crystalline
AMR has an additional dependence on the absolute orientation of ${\bf M}$ and
${\bf I}$ in the coordinate system of the crystal axes. We conclude the
qualitative discussion in Sec.~\ref{sec-II} by illustrating in the
Rashba-Dresselhaus system a potentially important effect on AMR of long-range
impurities, and in a spherical Kohn-Luttinger model\cite{Rushforth:2007_a} the
effect of carriers with higher spin state. Analysis of these effects relates
our work to previous theoretical studies of the AMR in (Ga,Mn)As diluted
magnetic semiconductors.\cite{Jungwirth:2002_c,Rushforth:2007_a,Rushforth:2007_b,Vyborny:2009_a}
The validity of the heuristic analysis of the AMR is confirmed in Sec.~\ref{sec-III} where
we explain the relation between the RTA and the exact solution to the integral
Boltzmann equation.\cite{Vyborny:2008_a} Quantitative results for the AMR are
derived in this Section and Appendix 
for the Rashba model and for the Dresselhaus model
with short-range electro-magnetic impurities and for the combined
Rashba-Dresselhaus model with arbitrary strength of the two SOI terms and with
short-range magnetic impurities. In Sec.~\ref{sec-IV} we comment on the
relevance of our model calculations to realistic two-dimensional semiconductor
structures.

\section{Heuristic link between spin textures and impurity potentials 
and the AMR}
\label{sec-II}

We limit our discussion in this section to AMRs defined as the relative
difference between longitudinal resistivities for magnetization aligned
parallel and perpendicular to the current direction. In situations discussed
below, the transverse resistivity vanishes and we can define
\begin{equation}\label{eq-03}
  \mbox{AMR} = \frac{\rho_{\hat{\bf I}}^{\parallel}-\rho_{\hat{\bf I}}^{\perp}}
               {(\rho_{\hat{\bf I}}^{\parallel}+\rho_{\hat{\bf I}}^{\perp})/2}=
           \frac{\sigma_{\hat{\bf I}}^{\perp}-\sigma_{\hat{\bf I}}^{\parallel}}
        {(\sigma_{\hat{\bf I}}^{\parallel}+\sigma_{\hat{\bf I}}^{\perp})/2}\,,
\end{equation}
where $\rho_{\hat{\bf I}}^{\parallel}$ ($\sigma_{\hat{\bf I}}^{\parallel}$)
and $\rho_{\hat{\bf I}}^{\perp}$($\sigma_{\hat{\bf I}}^{\perp}$) is the
longitudinal resistivity (conductivity) for ${\bf M}\parallel {\bf I}$ and for
${\bf M}\perp {\bf I}$, respectively, and the subscript $\hat{\bf I}$ labels
the orientation of current with respect to crystal axes. (The relation of our
microscopic theory to the standard phenomenology of the full angular
dependence of the AMR will be commented upon in Sec.~\ref{sec-III}.) Our
heuristic analysis of the AMR defined in Eq.~(\ref{eq-03}) is based on the RTA
and on assuming a proportionality between resistivity and the 1st order Born
approximation to elastic 
scattering probabilities from the state with the group velocity
along ${\bf I}$. Furthermore we consider only the strongest contribution to
the transport life-time which comes from back-scattering, i.e., from
transitions into states with group velocity opposite to ${\bf I}$. We use
these approximations and consider several archetypical spin-orbit coupled
Fermi surfaces to elucidate the relation of the spin-texture and nature of the
impurity potential to various fundamental aspects of the AMR phenomenology.

\subsection{AMR in the Rashba model}

We start with the two-dimensional electron system with Rashba SOI which yields
positive AMR independent of the current orientation in the crystal, and
demonstrate the potential qualitative difference between pure magnetic
short-range impurity potential $\propto \hat{\bf e}_M\cdot\hat{\bf s}/s$ and a
combined electro-magnetic potential 
$\propto\mathds{1}+\hat{\bf e}_M\cdot\hat{\bf s}/s$. 
Here $\hat{\bf e}_M$ denotes the magnetization unit
vector and $\hat{\bf s}$ is the carrier spin operator. For electrons with
$s=1/2$, the operator $\hat{\bf s}/s$ can be represented by the $2\times 2$
Pauli matrices~${\bg\sigma}=(\sigma_x,\sigma_y,\sigma_z)$.

\begin{figure}[h]
\includegraphics[width=0.7\columnwidth]{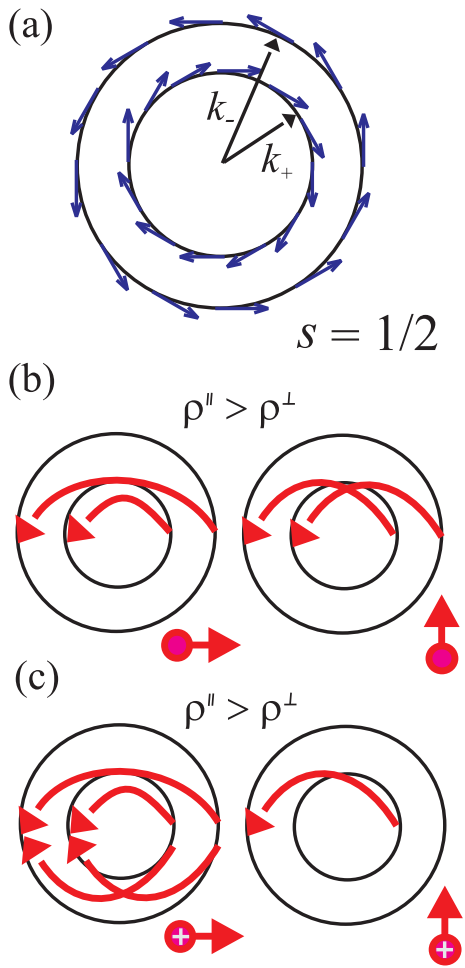}
\caption{Rashba model and (a) its spin texture along the Fermi
  contours. Dominant scattering channels for the states with group velocity
  pointing to the right when (b) magnetic and (c) electro-magnetic impurities
  (see text) constitute the prevalent source of momentum relaxation. Note the
  indicated directions of impurity polarization. The current flow
  is directed from left to right. The reader might
  consider the limit $k_-\gg k_+$ for better understanding of
  this and subsequent figures.}
\label{fig-01}
\end{figure}

\begin{figure}[h]
\includegraphics[width=0.8\columnwidth]{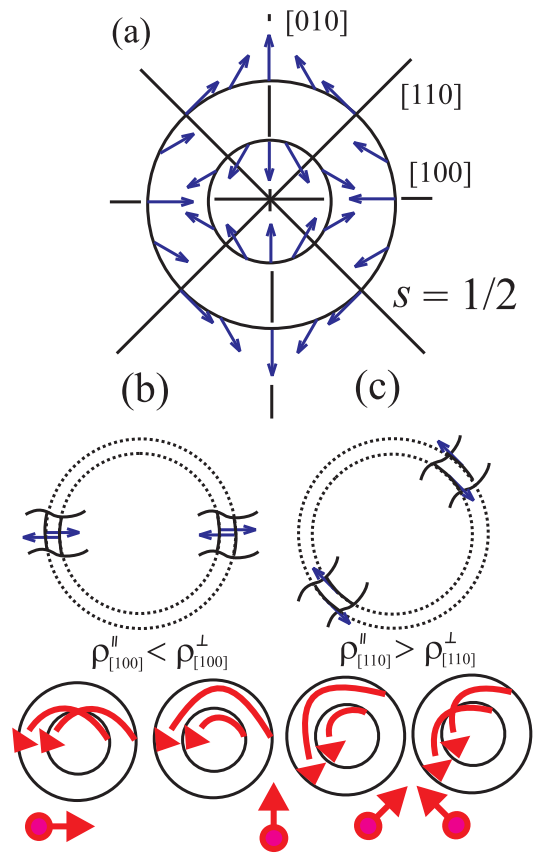}
\caption{Dresselhaus model and (a) its spin texture. In order to determine the
  current and the AMR along the $[100]$ and $[110]$ crystallographic
  directions we focus on the states with group velocities pointing in the
  respective directions, (b) and (c). Dominant momentum relaxation channels
  for these states and scattering on magnetic impurities are indicated on the
  bottom panels.}
\label{fig-02}
\end{figure}

The tangential spin-texture along the Fermi contour of the Rashba Hamiltonian,
\begin{equation}\label{eq-01a}
H_R=\frac{\hbar^2k^2}{2m}+\alpha(\sigma_x k_y-\sigma_y k_x)\,,
\end{equation}
is shown in Fig.~\ref{fig-01}(a). The spinors on the majority ($-$) 
and minority ($+$) Rashba band are given by $|\mathbf{k}_{\pm}\rangle = (1,\mp
ie^{i\theta})$, where $\tan\theta = k_y/k_x$.
From now on the coordinate system is chosen in such a way that
$\hat{x}$, $\hat{y}$, and $\hat{z}$ directions coincide
with [100], [010], and [001] crystallographic axes respectively,
as shown in Fig.~\ref{fig-02}.
Assuming current along
$\hat{x}$-direction, we can infer the back-scattering amplitudes of the states
with the group-velocity (${\mathbf k}$-vector) parallel to the current by
recalling the following properties of the scattering matrix elements:
\begin{equation}\label{eq-02}
  \begin{array}{lcl}
    \langle\downarrow|\sigma_x |\downarrow\rangle = 0 &&
    \langle\uparrow|\sigma_x |\downarrow\rangle= 1 \\
   \langle\downarrow|\sigma_y |\downarrow\rangle=1 &&
    \langle\uparrow|\sigma_y |\downarrow\rangle = 0\,.
  \end{array}
\end{equation}
Here we labeled the spinors by arrows whose orientation can be directly
compared to the spin-textures depicted in Fig.~\ref{fig-01}(a). The allowed
back-scattering processes, according to the relations in (\ref{eq-02}), are
highlighted in Fig.~\ref{fig-01}(b) for the pure magnetic impurity potential.
When magnetization points along the $\hat{x}$-direction
(i. e. to the right in Fig.~\ref{fig-01}(b)),
$\hat{e}_M\cdot{\bg\sigma}=\sigma_x$ and the back-scattering of states moving
along the $\hat{x}$-direction is due to majority-to-majority and
minority-to-minority band transitions. In the case of magnetization parallel
to the $\hat{y}$-direction, $\hat{e}_M\cdot{\bg\sigma}=\sigma_y$ and
back-scattering is due to majority-to-minority and minority-to-majority
transitions. In the limit of $k_-\gg k_+$, these figures suggest that
back-scattering is strongly suppressed for ${\bf M}\perp {\bf I}$ implying low
resistivity in this configuration compared to the ${\bf M}\parallel {\bf I}$
case. The AMR defined in Eq.~(\ref{eq-03}) is therefore expected to have
positive sign in the Rashba model. Quantitative Boltzmann equation
calculations presented in Sec.~\ref{sec-III} confirm the positive AMR for all
$k_->k_+$. They also confirm the vanishing magnitude of the AMR in the weak
SOI, large Fermi energy limit ($k_+\approx k_-$) which is discerned directly
from our pictorial representation of the allowed backs-scattering transitions
considering nearly degenerate majority and minority Rashba bands in
Fig.~\ref{fig-01}(b).

The behavior of AMR in the limit of degenerate Rashba bands, while keeping the
tangential spin textures, is qualitatively altered when the impurity potential
contains magnetic and non-magnetic components (e.g. for Mn acceptors in III-V
semiconductors). Replacing $\sigma_{x,y}$ with $\mathds{1}+\sigma_{x,y}$ in the
relations (\ref{eq-02}) allows us to illustrate this by again considering the
transitions that contribute to the back-scattering; note that this does 
{\em not} describe the situation where there are two distinct types of
impurities\footnote{The situation when two distinct types of impurities are present is treated in Ref. \onlinecite{Jungwirth:2002_c} and also discussed in Ref. \onlinecite{Vyborny:2009_a}.} (such as phonons and charge-neutral magnetic impurities). 
As highlighted in
Fig.~\ref{fig-01}(c), there is now always one of the Rashba bands in which
back-scattering is absent for ${\bf M}\perp {\bf I}$, independent of the
difference between $k_+$ and $k_-$. For ${\bf M}\parallel {\bf I}$,
back-scattering occurs in both bands and each of the states moving along the
current can scatter to both majority and minority band states. This implies
large positive AMR even in the limit of $k_+\approx k_-$.

Finally we point out that the circular symmetry of the Rashba spin-texture
makes the model a prototype realization of a purely non-crystalline AMR
system. The AMR is independent of the orientation of current in the coordinate
system of crystallographic axes and depends only on the relative angle between
${\bf M}$ and ${\bf I}$.

\subsection{AMR in the Dresselhaus model}
\label{sec-IIC}

The tangential spin-1/2 texture of the Rashba model represents arguably the
simplest host for a positive purely non-crystalline AMR. The Dresselhaus SOI
can be viewed as a minimal model demonstrating the link between a radial
spin-1/2 texture and a negative AMR, and illustrating the emergence of
crystalline AMR. The Dresselhaus Hamiltonian,
\begin{equation}\label{eq-01c}
H_D = \frac{\hbar^2k^2}{2m}+\beta(\sigma_x k_x-\sigma_y k_y)\;,
\end{equation}
yields the majority and minority eigenstates, $|\mathbf{k}_{\pm}\rangle=(1,\pm
e^{-i\theta})$, whose spin orientations along the respective Fermi contours
are depicted in Fig.~\ref{fig-02}(a). We can use the same analysis of the
back-scattering amplitudes as in the previous subsection to link this spin
texture to the expected basic AMR phenomenology in the Dresselhaus model.

In Fig.~\ref{fig-02}(b), we consider the case of current flowing along the
$\hat{x}$-direction ([100] crystal axis) and scattering from impurities
carrying the short-range magnetic potential only.  Using the same
representation of the spinors as in Eqs.~(\ref{eq-02}) we can write
\begin{equation}\label{eq-02_1}
  \begin{array}{lcl}
    \langle\rightarrow|\sigma_x |\rightarrow\rangle =1 &&
    \langle\leftarrow|\sigma_x |\rightarrow\rangle= 0 \\
   \langle\rightarrow|\sigma_y |\rightarrow\rangle= 0 &&
    \langle\leftarrow|\sigma_y |\rightarrow\rangle=1\;.
  \end{array}
\end{equation}
This implies that for magnetization parallel to the current direction,
back-scattering is due to majority-to-minority and minority-to-majority band
transitions while for magnetization perpendicular to the current, allowed
transitions are the majority-to-majority and minority-to-minority. The
low-resistivity and high-resistivity magnetization orientations therefore
switched places compared to the Rashba model and the AMR becomes negative.

The spin-texture of the Dresselhaus model is not circularly symmetric,
however. It evolves from radial for ${\bf k}$ parallel to the [100] or [010]
crystal axes to tangential for ${\bf k}$ parallel to the [110] or
[$\bar{1}$10] diagonals, as shown in Fig.~\ref{fig-02}(a). The back-scattering
amplitudes for current along the diagonal, highlighted in
Fig.~\ref{fig-02}(c), are hence identical as in the Rashba model, implying
positive AMR for this current direction. The lower symmetry of the Dresselhaus
model does not give rise to anisotropy in the resistivity of the system in the
absence of magnetization.\cite{Trushin:2006_a} 
However, when magnetization is present the system
acquires a crystalline AMR which reflects the underlying cubic symmetry of the
spin-texture.  We remark that both the negative and positive AMRs of the
Dresselhaus model vanish in the limit of $k_+\approx k_-$. Also in analogy
with the behavior of the Rashba model, the AMRs with the respective signs are
recovered in this limit when the pure magnetic impurity potential is replaced
with the combined electro-magnetic potential (see Sec.~\ref{sec-III} 
and Fig.~\ref{rashba_dress_el-mag_exact}).

\subsection{AMR in the Rashba-Dresselhaus model with $|\alpha|=|\beta|$}
\label{sec-IID}

\begin{figure}[h]
\includegraphics[width=0.7\columnwidth]{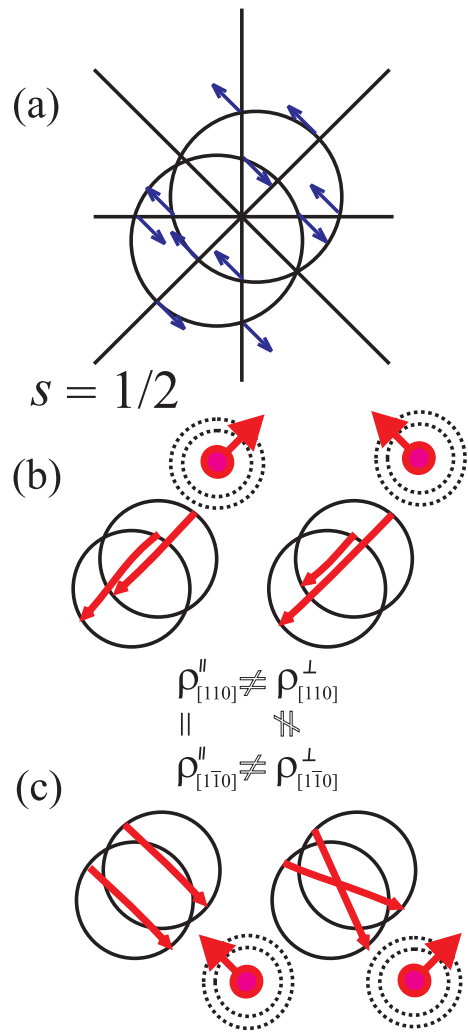}
\caption{(a) Spin texture along the Fermi contours of Rashba-Dresselhaus model
with $\alpha=\beta$. The AMR is zero for any type of short-range impurities.
However, for {\em long-range} magnetic impurities the scattering amplitudes
depend on the momentum transfer (illustrated by the length of the arrows)
and non-zero AMR arises for current both
along (b) $[110]$ and (c) $[\bar{1}10]$ crystallographic directions.  }
\label{fig-03}
\end{figure}

We now briefly comment on the potential importance of long-range nature of the
impurity potential on the basic AMR phenomenology. For the demonstration of
this effect, a singular model combining Rashba and Dresselhaus SOIs with
$|\alpha|=|\beta|$ is particularly suitable.  The Hamiltonian containing
Rashba and Dresselhaus spin-orbit coupling terms of equal strength has
singular properties\cite{Schliemann:2003_a,Schliemann:2003_b,Bernevig:2006_a}
(in particular additional symmetries).
The internal
spin-orbit coupling field has a ${\bf k}$-vector independent orientation
(along the $[1\bar{1}0]$-axis for $\alpha=\beta$). Spins on one circular Fermi
contour are aligned parallel to this field while on the other contour they
take the anti-parallel alignment. Additionally, as shown in
Fig.~\ref{fig-03}(a), this singular SOI shifts the two equal-size Fermi
contours with respect to each other along a direction perpendicular to the
direction of the spin-orbit field.

Because of the rigid spin-texture of the $|\alpha|=|\beta|$ Rashba-Dresselhaus
model on two mutually shifted but otherwise identical circular Fermi contours,
the back-scattering amplitudes for a short-range impurity potential are
independent of both the relative angle between ${\bf M}$ and the group
velocity of the state moving along ${\bf I}$, and of the direction of current
with respect to crystal axes.  The AMR therefore completely vanishes in this
model. Nevertheless, Figs.~\ref{fig-03}(b),(c) illustrate that the AMR,
including its crystalline component, is recovered when the scattering
amplitudes pick up a dependence on the transferred momentum, i.e., for
impurities carrying a long-range electro-magnetic potential.

\subsection{AMR in the spherical Kohn-Luttinger model}
\label{sec-IIE}

\begin{figure}[h]
\includegraphics[width=0.7\columnwidth]{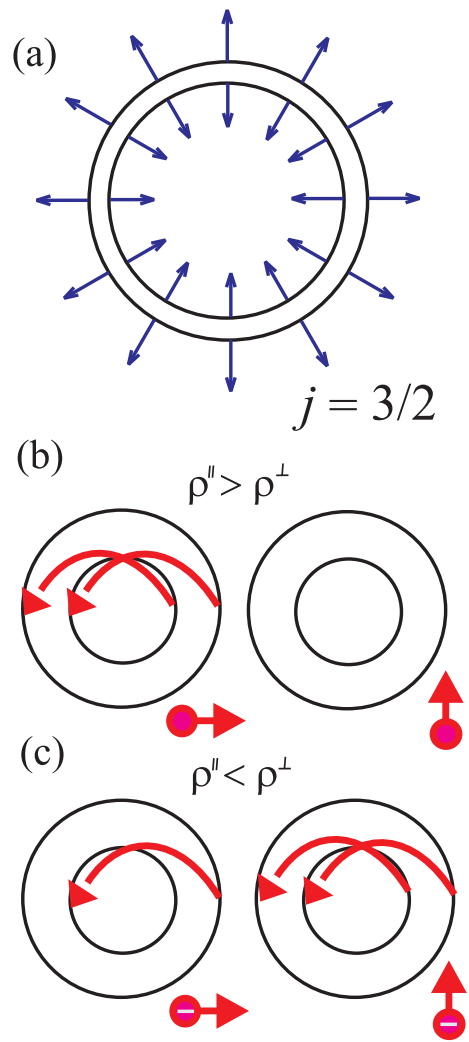}
\caption{(a) Cross-section (parallel to the $k_x,k_y$ plane) of the 3D 
  radial spin texture belonging to the two lower-energy bands of 
  the Kohn-Luttinger Hamiltonian. The two Fermi
  surfaces are sketched with different sizes for clarity, although the
  Hamiltonian~(\ref{eq-01d}) implies $k_-=k_+$ as $h\to 0$.
  (b) Dominant scattering channels for magnetic 
  impurities, note the difference to Fig.~\ref{fig-02}(b). (c) The same as (b)
  for electro-magnetic scatterers.}
\label{fig-kl}
\end{figure}

We conclude our excursion into the basic phenomenology of AMR, produced by
scattering of spin-orbit coupled carriers from polarized magnetic impurities,
by considering higher spin state of the carriers. We show that seemingly
identical spin-textures can result in opposite sign of the AMR for spin-1/2
and higher spin carriers, and argue that the AMR can have opposite sign when
carriers with higher spin are scattered from a pure magnetic or from a
combined electro-magnetic potential. Again seeking the minimal SOI model on
which this AMR phenomenology can be demonstrated without performing detailed
transport calculations we choose the four-band spherical 
{\em three-dimensional} Kohn-Luttinger Hamiltonian for total angular momentum
$j=3/2$ carriers,
\begin{equation}\label{eq-01d}
  H_{KL} = \frac{\hbar^2}{2m}\left[(\gamma_1+\frac52\gamma_2)k^2-
  2\gamma_2(\mathbf{k}\cdot\mathbf{j})^2\right]
  +hj_z
\end{equation}
with $h\to 0$.
The $k_x,k_y$ plane (with infinitesimal $k_z$) spin-textures depicted in Fig.~\ref{fig-kl}(a) are
obtained by realizing that the spin operator ${\bf s}={\bf j}/3$ in the
four-band model, by defining the momentum quantization axes parallel to 
${\bf k}$, and considering only the $j_{\bf k}=\pm 3/2$ bands (heavy holes).
The infinitesimal exchange field $h$ in Eq.~(\ref{eq-01d}) is included to lift 
the degeneracy of these two bands, and 
$\gamma_1$ and $\gamma_2$ are the Luttinger
parameters specific to the particular semiconductor valence bands for which 
$H_{KL}$ is derived from the conventional 
${\bf k}\cdot{\bf p}$ approximation.\cite{Sinova:2008_a,Vurgaftman:2001_a}

Unlike the spin-1/2 Dresselhaus model, the radial spin texture in the $j=3/2$
Kohn-Luttinger model yields a positive AMR for purely magnetic scatterers. 
This can be illustrated using an
analogous representation as in Eqs.~(\ref{eq-02_1}) to relate the scattering
amplitudes for impurity potential 
$\propto \hat{e}_{\bf M}\cdot{\bf s}/s=
\hat{e}_{\bf M}\cdot{\bf j}/j$ and the spin-texture. For the $j=3/2$
carriers we obtain\cite{Rushforth:2007_b}
\begin{equation}\label{eq-02_2}
  \begin{array}{lcl}
    \langle\rightarrow|j_x |\rightarrow\rangle \neq 0 &&
    \langle\leftarrow|j_x |\rightarrow\rangle= 0 \\
   \langle\rightarrow|j_y |\rightarrow\rangle= 0 &&
    \langle\leftarrow|j_y |\rightarrow\rangle= 0\;.
  \end{array}
\end{equation}
This implies, as highlighted in Fig.~\ref{fig-kl}(b), that for magnetization
parallel to the current direction, back-scattering is due to
majority-to-minority and minority-to-majority band transitions as in the case
of spin-1/2 carriers. However, for magnetization perpendicular to the current,
there are no allowed back-scattering transitions in contrast to the spin-1/2
Dresselhaus model in Fig.~\ref{fig-02}(b). 
This makes now the latter configuration the low-resistivity
state and AMR for the radial spin-texture of the Kohn-Luttinger model becomes
positive for pure magnetic impurity potential even for $k_+\approx k_-$. 
Boltzmann equation calculation of the AMR presented in Appendix~\ref{app-D}
(and also an independent calculation based on the Green's function
formalism\cite{Kovalev:2009_a}) 
again confirm our heuristic conclusion of Fig.~\ref{fig-kl}(b). 

On the other hand, electro-magnetic scatterers 
$\propto \frac{3}{2}\mathds{1}+j_{x,y}$
produce negative AMR in the Kohn-Luttinger 
model\cite{Rushforth:2007_b,Vyborny:2009_a} in the
very same way as it is shown in Fig.~\ref{rashba_dress_el-mag_exact}(b) for
the Dresselhaus model, and in both cases, this behavior can again be inferred
using relations~(\ref{eq-02_2})~and~(\ref{eq-02_1}) with $j_{x,y}$ and
$\sigma_{x,y}$ replaced by $\frac{3}{2}\mathds{1}+j_{x,y}$ and
$1+\sigma_{x,y}$, respectively. 
Dominant scattering channels which suggest that $\mbox{AMR}<0$ 
are summarized in Fig.~\ref{fig-kl}(c).
Contrary to the Dresselhaus
model~(\ref{eq-01c}), the SOI of the Kohn-Luttinger model~(\ref{eq-01d})
in combination with polarized scatterers therefore can produce AMR of either
sign, e.g. depending on the carrier-density-controlled screening of the
impurities.\cite{Vyborny:2009_a,Rushforth:2007_b} 
This qualitative difference between
Dresselhaus and Kohn-Luttinger models highlights the fact that
knowledge of spin textures, such as Figs.~\ref{fig-02}(a) or~\ref{fig-kl}(a),
may not be sufficient to analyze the scattering properties of the model and
appropriate matrix elements such as Eqs.~(\ref{eq-02_1}) or~(\ref{eq-02_2})
should always be verified.

\section{Quantitative results for the AMR in the Rashba-Dresselhaus model}
\label{sec-III}

The AMR analysis in the previous Section utilizes the RTA (in fact only the
back-scattering term of the RTA) which, in general, is not a rigorous theory
approach for anisotropic systems.\cite{Vyborny:2008_a} It is therefore
desirable to calculate the AMR beyond the RTA, not only to obtain quantitative
predictions but also to confirm the validity of the basic AMR phenomenology
inferred above. As in Sec.~\ref{sec-II}, we will employ the 1st order Born
approximation for calculating the scattering probabilities but will solve the
corresponding integral Boltzmann equation exactly.  To provide better physical
insight we start with explaining the relation between the RTA and the full
semiclassical Boltzmann theory for the two-dimensional SOI systems.
Exact analytical solutions to the Boltzmann equation are
then derived for Rashba and Dresselhaus model with short range
electro-magnetic impurity potentials and for the combined Rashba-Dresselhaus
model with arbitrary $\alpha$ and $\beta$ and with magnetic impurities.

\subsection{Relation between RTA and integral Boltzmann equation
in the Rashba model}
\label{sec-IIIA}

Because the equilibrium Fermi distribution $f^0(E_{i,\mathbf{k}})$ is a
function only of energy, we can write the Boltzmann
equation\cite{Vyborny:2008_a} in $d=2$ dimensions as
\begin{widetext}
\begin{eqnarray}\label{eq-05}
%\label{master2}
-|e|\mathbf{E}\cdot\mathbf{v}_{i,\mathbf{k}}
\frac{\partial f^0(E_{i,\mathbf{k}})}{\partial E_{i,\mathbf{k}} }&=&
-\int\frac{d^d k'}{(2\pi)^d}\sum_{i'}
w(i,\mathbf{k};i',\mathbf{k'})\delta(E_{i',\mathbf{k}'}-E_{i,\mathbf{k}})
\left[f(i,\mathbf{k})-f(i',\mathbf{k}')\right]\nonumber \\
&=&-\left[f(i,\mathbf{k})-f^0(E_{i,\mathbf{k}})\right]\sum_{i'}\int\frac{d^dk'}{(2\pi)^d}
w(i,\mathbf{k};i',\mathbf{k'})\delta(E_{i',\mathbf{k}'}-E_{i,\mathbf{k}})\nonumber \\
&+&\int\frac{d^dk'}{(2\pi)^d}\sum_{i'}
w(i,\mathbf{k};i',\mathbf{k'})\delta(E_{i',\mathbf{k}'}-E_{i,\mathbf{k}})
\left[f(i',\mathbf{k}')-f^0(E_{i',\mathbf{k}'})\right]\;,
\end{eqnarray}
\end{widetext}
where
$\mathbf{v}_{i,\mathbf{k}}=\partial E_{i,\mathbf{k}}/\partial \hbar\mathbf{k}$
is the group velocity, $f(i,\mathbf{k})$ is the non-equilibrium distribution
function, and $i=\pm$ is the band index. The transition probabilities in the
1st order Born approximation are given by
\begin{equation}
w(i,\mathbf{k};i',\mathbf{k}')=
\frac{2\pi n}{\hbar}|\langle i,\mathbf{k}|V|i',\mathbf{k}'\rangle|^2
\;,
\label{w}
\end{equation}
where $V$ is the strength of the short-range scattering potential of
impurities with density $n$. Energy conservation during elastic scattering
processes was already incorporated into the right hand-side of 
Eq.~(\ref{eq-05}).

In the Rashba model, $\sum_{i'}w(i,\mathbf{k};i',\mathbf{k}')$ is a
constant\footnote{See Eq.~(\ref{wRashba}) and Tab.~\ref{tab-01} for
  explicit expressions in the case of $V\propto
  \hat{e}_M\cdot{\bg\sigma}$. For $V\propto \mathds{1}$, we find $\Sigma_{i'}
  w(i,\theta;i',\theta)$ independent of $\theta,\theta'$ using
  \hbox{$w(i,\theta;i',\theta')=1+ii'\cos(\theta-\theta')$.}}
for a short-range electric potential, $V\propto \mathds{1}$, or magnetic
potential, $V\propto \hat{e}_M\cdot{\bg\sigma}$. In the limit of nearly
degenerate bands, $E_{i,\mathbf{k}}\approx E_{i',\mathbf{k}}$, we can find a
solution of Eq.~(\ref{eq-05}) in the RTA form,
\begin{equation}
f(i,\mathbf{k})-f^0(E_{i,\mathbf{k}})=c|e|\mathbf{E}\cdot\mathbf{v}_{i,\mathbf{k}}
\frac{\partial f^0(E_{i,\mathbf{k}})}{\partial E_{i,\mathbf{k}} }\;.
\label{rta}
\end{equation}
Plugged in Eq.~(\ref{eq-05}), the second term on the right-hand side drops out
because of the independence of $\sum_{i'}w(i,\mathbf{k};i',\mathbf{k}')$ on
$\mathbf{k}'$ and because the group velocity averages to zero over the Fermi
contour, and the first term gives
\begin{equation}
\frac{1}{c}=\int\frac{d^2k'}{(2\pi)^2}\sum_{i'}
w(i,\mathbf{k};i',\mathbf{k'})\delta(E_{i',\mathbf{k}'}-E_{i,\mathbf{k}})
\equiv\frac{1}{\tau}\;.
\label{tau}
\end{equation}
The electrical current within the semiclassical linear response, given by
\begin{equation}
\label{current}
\mathbf{j}=-e\sum_i\int\frac{d^2 k}{(2\pi)^2}
   \mathbf{v}_{i,\mathbf{k}} \left[f(i,\mathbf{k})-f^0(i,\mathbf{k})\right]
\;,
\end{equation}
is exactly proportional to the quasiparticle broadening life-time $\tau$ in
this case. Same RTA form of the Boltzmann equation applies also to the
Rashba-Dresselhaus model with $|\alpha|=|\beta|$ because the rigid
spin-texture of this singular case implies constant transition probabilities
for any short-range electro-magnetic potential.

In the Rashba model with non-degenerate bands, $E_{i,\mathbf{k}}\neq
E_{i',\mathbf{k}}$, the RTA solution (\ref{rta}) to the Boltzmann equation can
still be found for a non-magnetic potential, $V\propto \mathds{1}$. The
scattering probability $w(i,\mathbf{k};i',\mathbf{k}')$ depends in this case
on the magnitude of the transition angle, $|\theta-\theta'|$. It implies that
from the product,
\begin{equation}
\mathbf{E}\cdot\mathbf{v}_{i',\mathbf{k}'}=v_i\mathbf{E}\cdot\mathbf{v}_{\mathbf{k}}\frac{v_{i'}}{v_i}
\cos(\theta-\theta')+v_i(\hat{\mathbf{z}}\times\mathbf{E})\cdot\mathbf{v}_{\mathbf{k}}\frac{v_{i'}}{v_i}
\sin(\theta-\theta')\;,
\label{Ev}
\end{equation}
the transverse term $\propto\sin(\theta-\theta')$ does not contribute to the
second term on the right-hand side of Eq.~(\ref{eq-05}). The longitudinal term
$\propto\cos(\theta-\theta')$ contributes to Eq.~(\ref{eq-05}) and the
Boltzmann equation takes a modified RTA form with
\begin{eqnarray}\nonumber
\frac{1}{c}&=&\int\frac{d^2k'}{(2\pi)^2}\sum_{i'}
w(i,\mathbf{k};i',\mathbf{k'})\delta(E_{i',\mathbf{k}'}-E_{i,\mathbf{k}})\\
&\times&\left[1-\frac{v_{i'}}{v_i}\cos(\theta-\theta')\right]
\equiv\frac{1}{\tau_{tr}}\;.
\label{tautr}
\end{eqnarray}
Electrical current is now proportional to the transport life-time which gives
larger weight to larger angle scattering transitions.

The transport life-time form of the Boltzmann equation has been the basis of
qualitative discussions in Sec.~\ref{sec-II} where we further simplified the
analysis by considering only the leading contribution to current in
Eq.~(\ref{current}) from states with
$\mathbf{v}_{i,\mathbf{k}}\parallel\mathbf{E}$. For all spin-textures and
orientations of $\mathbf{E}$ and $\mathbf{M}$ considered in Sec.~\ref{sec-II},
$w(i,\mathbf{k};i',\mathbf{k}')$ depends only on $|\theta-\theta'|$ for the
{\em special $\mathbf{k}$-states} with group velocity parallel to the electric
field. This justifies the internal consistency of the RTA based analyses in
Sec.~\ref{sec-II} and explains their qualitative validity.

\subsection{Solution to the Boltzmann equation for the Rashba-Dresselhaus 
model}

To obtain quantitative AMR predictions we need to perform the full
$\mathbf{k}$-space integration in the expression (\ref{current}) for the
electrical current. For arbitrary $\mathbf{k}$-vector and other than the few
special cases discussed in the previous subsection (which all happen to give
zero AMR), the integral of the transverse term in Eq.~(\ref{Ev}) may not
vanish and/or the integrated scattering probability in the first term on the
right-hand side of Eq.~(\ref{eq-05}) may not be independent of ${\mathbf k}$.
In these cases the RTA form of the solution to the Boltzmann equation fails.
For the Rashba-Dresselhaus model we can, nevertheless, find the exact solution
to the Boltzmann equation in an analytic form which allows us to directly
compare the corresponding quantitative AMR predictions with the qualitative
results of Sec.~\ref{sec-II}.

The method has been previously derived\cite{Vyborny:2008_a} for pure Rashba
model in which the angular dependence of the scattering probability function
for the short range magnetic potential, e.g. $V\propto \sigma_x$, is given by
\begin{equation}\label{wRashba}
w(i,\theta;i',\theta')\propto 
1-ii'(\cos\theta\cos\theta'-\sin\theta\sin\theta')\;.
\end{equation}
Since also $\int_0^{2\pi} d\theta' w(i,\theta;i',\theta')$ is a constant
independent of $\theta$, the first term on the right-hand side of
Eq.~(\ref{eq-05}) implies that $f(i,\mathbf{k})-f^0(E_{i,\mathbf{k}})$ must
contain term 
$\mathbf{E}\cdot{\mathbf v}_{i,k}(\theta)$ and the second term on the
right-hand side of Eq.~(\ref{eq-05}) implies that
$f(i,\mathbf{k})-f^0(E_{i,\mathbf{k}})$ must contain harmonics of
$w(i,\theta;i',\theta')$ which in both cases happen to be just $\cos\theta$
and $\sin\theta$. No higher order Fourier components can contribute to the
non-equilibrium distribution function in this case and Eq.~(\ref{eq-05}) can
be solved analytically.

\begin{figure}[h]
\begin{center}\begin{tabular}{cc}
\includegraphics[height=4.5cm]{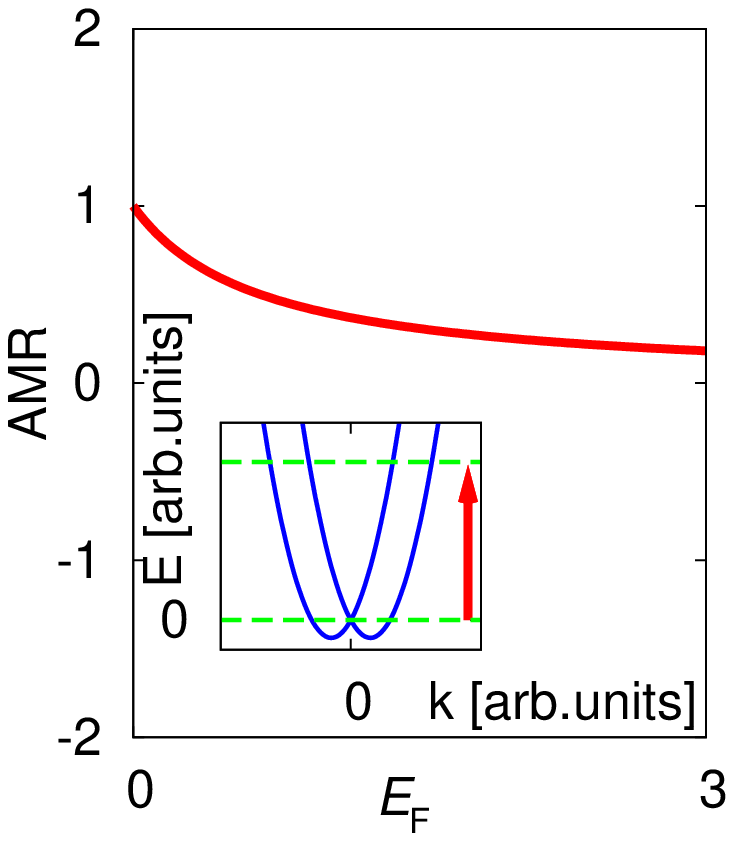}&
\hskip-3cm\includegraphics[height=4cm]{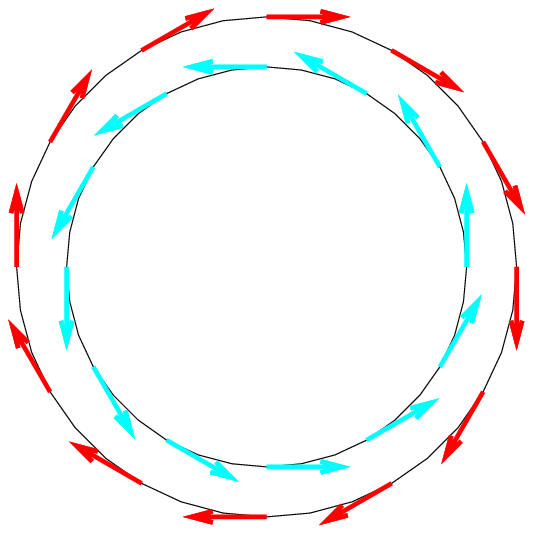} 
\end{tabular}\end{center}
\caption{Pure Rashba system with magnetic impurity, AMR as a function 
of the Fermi energy $E_F$ in units of $m \alpha^2/\hbar^2$.}
\label{rashba_mag_exact}
\end{figure}

The AMR of the Rashba model with magnetic impurity potential is summarized in
the first column of Tab.~\ref{tab-02} and also plotted in Fig.~\ref{rashba_mag_exact} as a
function of the ratio $E_F\hbar^2/(m\alpha^2)$. Here $E_F=0$ corresponds to
the minority Rashba band being just depleted and $E_F\hbar^2/(m\alpha^2)\gg 1$
to nearly degenerate $i=\pm$ Rashba bands. Consistent with the qualitative
results of Sec.~\ref{sec-II} we find a positive AMR which vanishes as the
radii of the minority and majority band Fermi contours approach each other.

\begin{table*}
\begin{tabular}{|c|c|c|c|}
\hline
Magnetization & $\alpha\neq 0$ & $\alpha=0$ & $\alpha=\beta$ \\
of scatterers &  $\beta=0$ &  $\beta\neq 0$ & \\ \hline
Along [100]  & $\hat{\sigma}=\left(\begin{array}{cc} \sigma_0-\frac{2}{3}A & 0 \\  0 & \sigma_0 \end{array}\right)$ &  $\hat{\sigma}=\left(\begin{array}{cc} \sigma_0 & 0 \\  0 & \sigma_0-\frac{2}{3}B \end{array}\right)$ &  $\hat{\sigma}=\left(\begin{array}{cc} \sigma_0 & 0 \\  0 & \sigma_0 \end{array}\right)$ \\ \hline
Along [010]  & $\hat{\sigma}=\left(\begin{array}{cc} \sigma_0 & 0 \\  0 & \sigma_0-\frac{2}{3}A \end{array}\right)$  &  $\hat{\sigma}=\left(\begin{array}{cc} \sigma_0-\frac{2}{3}B & 0 \\  0 & \sigma_0 \end{array}\right)$ & $\hat{\sigma}=\left(\begin{array}{cc} \sigma_0 & 0 \\  0 & \sigma_0 \end{array}\right)$ \\ \hline
Along [110] & $\hat{\sigma}=\left(\begin{array}{cc} \sigma_0-\frac{1}{3}A & -\frac{1}{3}A \\  -\frac{1}{3}A& \sigma_0-\frac{1}{3}A \end{array}\right)$  & $\hat{\sigma}=\left(\begin{array}{cc} \sigma_0-\frac{1}{3}B & -\frac{1}{3}B \\  -\frac{1}{3}B & \sigma_0-\frac{1}{3}B \end{array}\right)$  & $\hat{\sigma}=\left(\begin{array}{cc} \sigma_0 & 0 \\  0 & \sigma_0 \end{array}\right)$  \\ \hline
Along [1$\bar{1}$0] & $\hat{\sigma}=\left(\begin{array}{cc} \sigma_0-\frac{1}{3}A & \frac{1}{3}A \\  \frac{1}{3}A& \sigma_0-\frac{1}{3}A \end{array}\right)$  & $\hat{\sigma}=\left(\begin{array}{cc} \sigma_0-\frac{1}{3}B & \frac{1}{3}B \\  \frac{1}{3}B & \sigma_0-\frac{1}{3}B \end{array}\right)$  & $\hat{\sigma}=\left(\begin{array}{cc} \sigma_0 & 0 \\  0 & \sigma_0 \end{array}\right)$
%Along [001] & $\sigma=\left(\begin{array}{cc} \sigma_0-\frac{2}{3}A & 0 \\  0 & \sigma_0-\frac{2}{3}A \end{array}\right)$  & $\sigma=\left(\begin{array}{cc} \sigma_0-\frac{2}{3}B & 0 \\  0 & \sigma_0-\frac{2}{3}B \end{array}\right)$  & $\sigma=\left(\begin{array}{cc} \sigma_0 & 0 \\  0 & \sigma_0 \end{array}\right)$
\\ \hline
\end{tabular}
\caption{Conductivity tensor for a 2DEG confined in a
$[001]$-grown III-V semiconductor heterostructure at different
magnetization of scatterers. Here,  $\sigma_0=e^2 n_e \tau/m$, see also
Eq.~(\ref{eq-30}),  $n_e$ is the electron density, see Eq.~(\ref{eq-17}), 
and $A/\alpha^2=B/\beta^2=e^2 m \tau/(\pi \hbar^4)$.
The conductivity corrections depend essentially on
the type of spin-orbit interactions which is either
Rashba ($\alpha$) or Dresselhaus ($\beta$) one.
The conductivity expressions for arbitrary $\alpha$ and $\beta$ can be
found in Appendix~\ref{app-B}.}
%\label{tab1}
\label{tab-02}
\end{table*}

\begin{figure}[h]
%\includegr..cs[width=0.9\columnwidth,angle=90]{rashba_dress_el-mag_exact.eps}
\begin{center}\begin{tabular}{cc}
\includegraphics[height=4cm]{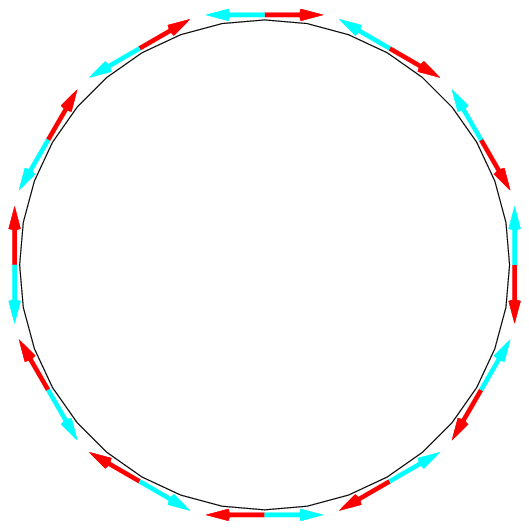}&%
\hskip-2cm\includegraphics[height=4cm]{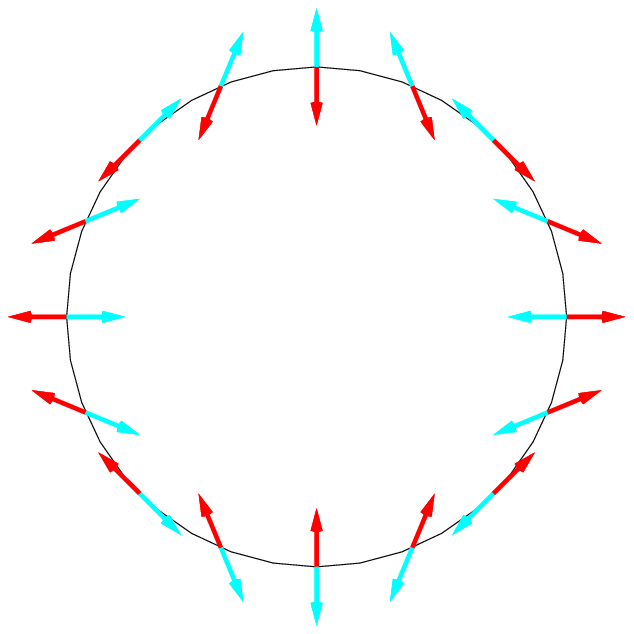}
\\[-1cm]
\includegraphics[height=4.5cm]{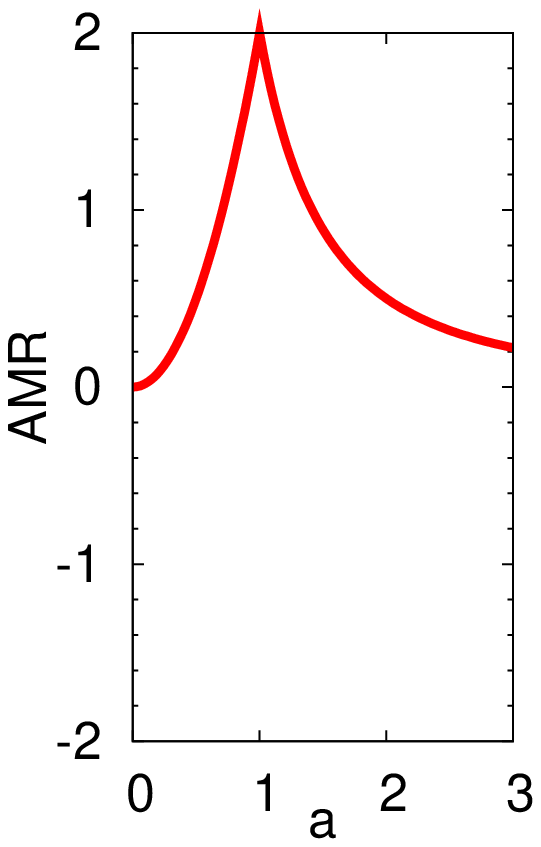}&%
\hskip-2.5cm\includegraphics[height=4.5cm]{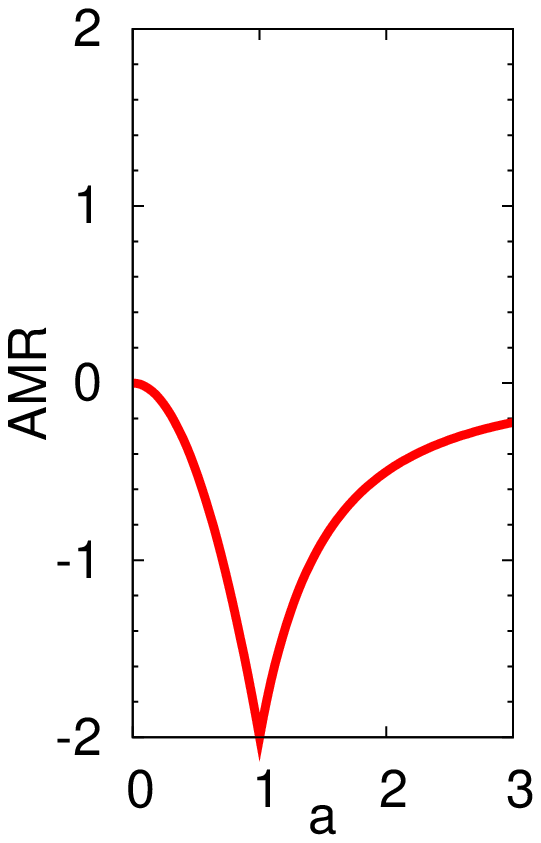}\\[-1cm]
\hskip-4cm\Large(a) & \hskip-6cm\Large(b)
\end{tabular}\end{center}
\caption{AMR for current flowing along $[100]$ crystal axis in a 
pure (a) Rashba and (b) Dresselhaus systems with {\em electromagnetic}
  impurity ($\propto a\mathds{1}+\sigma_{x,y}$), varying $a$, 
  the ratio between the electric
  and magnetic part of the potential. The Fermi energy $E_F$ is taken much
  larger than the spin-orbit interaction, so that the Fermi radii of the two
  bands become almost equal, see details in text.}
\label{rashba_dress_el-mag_exact}
\end{figure}

For Rashba model with the electro-magnetic potential, e.g. $V\propto
\mathds{1}+\sigma_x$, the integral $\int_0^{2\pi} d\theta'
w(i,\theta;i',\theta')\propto 1+ii'\sin\theta$ is not a constant which implies
the presence of higher order Fourier components in
$f(i,\mathbf{k})-f^0(E_{i,\mathbf{k}})$. Still an analytical form can be found
for the distribution function, see the note added in proof of
Ref.~\onlinecite{Vyborny:2008_a}. Analogous arguments apply
also to the Dresselhaus model with electro-magnetic impurities. The dependence
of AMRs in the two models as a function of the ratio $a$ of the electrical and
magnetic parts of the impurity potential $V\propto
a\mathds{1}+\hat{e}_M\cdot{\bg\sigma}$ in the limit of nearly degenerate bands
and for current along the [100]-axis is given by
\begin{equation}
\mbox{AMR}=\left\{ \begin{array}{ll} \pm 2a^2, & \mbox{for }|a|\le 1\\
                                       \pm 2/a^2,&\mbox{for }|a|\ge 1\;,
                                     \end{array}\right.
\end{equation}
where +/$-$ corresponds to the Rashba/Dresselhaus model. For illustration, we
also plot the result in Fig.~\ref{rashba_dress_el-mag_exact}.  Again in full
qualitative agreement with the analysis in Sec~\ref{sec-II}, the AMRs in both
models are zero for $a=0$. They also vanish in the limit of
$a\rightarrow\infty$ since no AMR occurs if the system is not magnetically
polarized. For intermediate ratios of the strengths of the electric and
magnetic parts of the potential, a positive AMR in the Rashba model reflects
the tangential spin-1/2 texture while the negative AMR in the Dresselhaus
model reflects the radial texture of the states with large group velocity
projection to the direction of the current. The singular peak at $a=1$
originates from the coherent superposition of non-magnetic and magnetic
scattering amplitudes which results in zero scattering probability of one of
the two states moving along the current direction,\footnote{See also discussion of physical relevance of this singularity in Sec. IV or Ref.~\onlinecite{Vyborny:2008_a}.} 
as we already pointed out in Sec~\ref{sec-II} and illustrated in Fig.~\ref{fig-01}(c).

In Tab.~\ref{tab-02}, we included conductivity components obtained from the
exact solution to the Boltzmann equation for Rashba and Dresselhaus models and
the magnetic potential with $\mathbf M$ oriented along the main in-plane
crystal axes and along the in-plane diagonals (derived as shown below). 
The component $\sigma_{11}$ in
the table corresponds to the longitudinal response to ${\mathbf E}$ along the
[100]-axis and $\sigma_{22}$ along the [010]-axis. To obtain AMR values for
electric field along an arbitrary angle $\phi$ measured from the [100]-axis
the conductivity tensors with appropriate magnetization direction of
scatterers have to be rotated by $R_{-\phi}\hat\sigma R_{\phi}$
where the rotation matrix is given by,
\begin{equation}
R_\phi=\left(\begin{array}{cc}
     \cos\phi & -\sin\phi \\
     \sin\phi & \cos\phi
               \end{array}\right)\;.
\end{equation}
The AMR as defined in Eq.~(\ref{eq-03}) is independent of $\phi$ in the Rashba
model confirming the absence of crystalline AMR components in this system. In
the Dresselhaus model, AMRs of opposite sign are obtained for current along
the main in-plane axes ($\phi=0,\pi/2$) and along the diagonals
($\phi=\pi/4,3\pi/4$), consistent with the crystalline nature of the AMR
inferred in Sec~\ref{sec-II}. A closer inspection of the full angular
dependence of the AMR in the Rashba and Dresselhaus models allows us to relate
our quantitative microscopic results to the standard phenomenology of the
angle-dependent longitudinal resistivity for systems with cubic
anisotropies,\cite{Rushforth:2007_a}
\begin{eqnarray}\label{eq-35}
  \rho(\omega,\phi)/\rho_{av} - 1 &=& C_{I}\cos 2(\omega-\phi) + \\ \nonumber
  &&C_{I,c}\cos 2(\omega+\phi) + C_c\cos 4\omega \;,
\end{eqnarray}
where $\omega$ and $\phi$ denote the direction angles of ${\mathbf M}$ and
${\mathbf E}$ to the $[100]$ crystal axis, respectively, and $\rho_{av}$ is
the average resistivity over all magnetization directions.  The coefficient
$C_I$ of the non-crystalline AMR component, which depends only on the relative
angle between current and magnetization, equals 1/3 for the Rashba model and 0
for the Dresselhaus model. The coefficient $C_{I,c}$ of the first crystalline
component is non-zero (equals $-1/3$) in the Dresselhaus model and zero in the
Rashba model, consistent with the crystalline nature of the AMR in the
Dresselhaus SOI system and non-crystalline AMR of the Rashba system. The
coefficient $C_c$ of the higher order crystalline term is zero in both models.

\begin{table}
\begin{tabular}{ccl}
$\hat{e}_M$ &\hbox to2mm{\hfill}& $P^{\hat{e}_M}(i,\theta;i',\theta')$ \\ 
\hline
$[100]$ && $1+ii'\cos(\gamma_k+\gamma_{k'})$ \\
$[010]$ && $1-ii'\cos(\gamma_k+\gamma_{k'})$ \\
$[110]$ && $1-ii'\sin(\gamma_k+\gamma_{k'})$ \\
$[1\bar{1}0]$ && $1+ii'\sin(\gamma_k+\gamma_{k'})$ \\
$[001]$ && $1-ii'\cos(\gamma_k-\gamma_{k'})$
\end{tabular}
\caption{Magnetization-direction-dependent factors $P^{\hat{e}_M}$ of
Eq.~(\ref{eq-10}) relevant for magnetic impurities.
Functions $\cos\gamma_k$, $\sin\gamma_k$ are given in the main text.}
\label{tab-01}
\end{table}

We conclude this Section by presenting the exact solution to the Boltzmann
equation and the corresponding AMR values for the combined Rashba-Dresselhaus
model which implies the dispersion law
$E_{\pm,\mathbf{k}}=\frac{\hbar^2 k^2}{2m}\pm k\kappa_\theta$,
where 
$\kappa_\theta=\sqrt{\alpha^2+\beta^2+2\alpha\beta\sin2\theta}$
is the $\theta$-dependent subband spin splitting.
We consider a general case of 
arbitrary $\alpha$ and $\beta$ but restrict ourselves to the pure magnetic impurity
potential.
The electron group velocity $(1/\hbar)\nabla_{\mathbf{k}} E_{\mathbf{k}\pm}$
is now anisotropic and given by
\begin{eqnarray}\label{eq-15a}
%\label{vx}
&&
\mathbf{v}_{\pm,k}\vert_x=
\hbar k_x/m \pm (\beta\cos\gamma_k + \alpha\sin\gamma_k)/\hbar\,, \\
%\label{vy}
\label{eq-15b}
&&
\mathbf{v}_{\pm,k}\vert_y=
\hbar k_y/m \pm (\alpha\cos\gamma_k + \beta\sin\gamma_k)/\hbar
\end{eqnarray}
with $k_x=k\cos\theta$, $k_y=k\sin\theta$, and
$\sin\gamma_k = (\alpha\cos\theta + \beta\sin\theta)/\kappa_{\theta}$,
$\cos\gamma_k = (\beta\cos\theta + \alpha\sin\theta)/\kappa_{\theta}$.

The derivation relies on vanishing angular integrals of the
generating functions of $w(i,\mathbf{k};i',\mathbf{k}')\propto
P(i,\theta;i',\theta')$ (summarized in Tab.~\ref{tab-01} for $\mathbf{M}$
along the main in-plane crystal axes and the in-plane diagonals) which are 
$\cos\theta/\kappa_{\theta}$ and $\sin\theta/\kappa_{\theta}$.
As in the case of the Rashba model discussed above, the independence of 
$\int_0^{2\pi} d\theta' w(i,\theta;i',\theta')$ on $\theta$ implies 
that the non-equilibrium distribution function contains only the group
velocity, see Eqs.~(\ref{eq-15a},\ref{eq-15b}), and the generating functions
of $P(i,\theta;i',\theta')$ which are
$\cos\theta/\kappa_\theta$ and $\sin\theta/\kappa_\theta$. Note
that for arbitrary $\alpha$ and $\beta$ and for the orientations of
$\mathbf{M}$ considered in Tabs.~\ref{tab-02} and \ref{tab-01} the transition
probabilities can then be written as,
\begin{equation}\label{eq-10}
  w(i,\mathbf{k};i'\mathbf{k}')=\frac{1}{\nu\tau}
  P^{\hat{e}_M}(i,\theta;i'\theta')\;,
\end{equation}
where $\nu=m/\pi\hbar^2$ is the density of states, the ${\mathbf k}$-vector
independent constant $\tau$ is given by Eq.~(\ref{tau}), and the angular
probabilities $P^{\hat{e}_M}(i,\theta;i'\theta')$ are explicitly written in
Tab.~\ref{tab-01}. 
The integral Boltzmann equation~(\ref{eq-05}) is then solved by the
distribution function of a form 
\begin{widetext}
\begin{eqnarray}\label{eq-08a}
%\label{solution1}
f(i,\mathbf{k})-f^0(E_{i,\mathbf{k}})=& &\tau |e|\mathbf{E}\cdot\mathbf{v}_{i,\mathbf{k}}
\frac{\partial f^0(E_{i,\mathbf{k}})}{\partial E_{i,\mathbf {k}} }\nonumber \\
&+&\frac{\tau |e|}{\hbar}
\frac{\partial f^0(E_{i,\mathbf{k}})}{\partial E_{i,\mathbf{k}} }
\left[\bigg(a_x^{\hat{e}_M}\frac{\cos\theta}{\kappa_\theta}
+b_x^{\hat{e}_M}\frac{\sin\theta}{\kappa_\theta}\bigg)E_x+
\bigg(a_y^{\hat{e}_M}\frac{\cos\theta}{\kappa_\theta}+
b_y^{\hat{e}_M}\frac{\sin\theta}{\kappa_\theta}\bigg)E_y\right]\;.
\end{eqnarray}
\end{widetext}
Values of the coefficients $a^{\hat{e}_M}_{x,y}$, $b^{\hat{e}_M}_{x,y}$ depend
on the magnetization vector direction $\hat{e}_M$ and are given in
Appendix~\ref{app-B}.

For $|\alpha|=|\beta|$, and general $\alpha$, $\beta$, analytical expressions
for the conductivity tensor of the Rashba-Dresselhaus model and short-range
magnetic impurity potential with ${\mathbf M}$ oriented along the main and
diagonal in-plane axes can be found in Tab.~\ref{tab-02}, and
Tab.~\ref{tab-03} in Appendix, respectively. As pointed out in
Sec.~\ref{sec-II}, the AMR vanishes for $|\alpha|=|\beta|$. For
$\alpha\neq\beta$, however, the AMR is non-zero and depends both on the
relative angle between current and magnetization and on the direction of
current with respect to the crystallographic axes. The AMRs for various
current directions can again be calculated by rotating the conductivity tensor
given in Tab.~\ref{tab-03}. For current along the [100]-axis, e.g., and
$|\alpha|\ge|\beta|$ we obtain
\begin{equation}\label{eq-31}
  \mbox{AMR} = \frac{2(1-r^2)^2}{2(1+r^2)^2+(3+r^2)\hbar^2 E_F/(m\alpha^2)}\;,
\end{equation}
where $r=\beta/\alpha$. In the opposite case of $|\alpha|\le |\beta|$, the
result is the same up to an exchange of $\alpha$ and $\beta$ in
Eq.~(\ref{eq-31}) and in the definition of $r$.

\begin{figure}
\begin{tabular}{cc}
%\hskip-.5cm\includegraphics[height=3.7cm]{fig2a.eps} &
\hskip-.5cm\includegraphics[height=3.7cm]{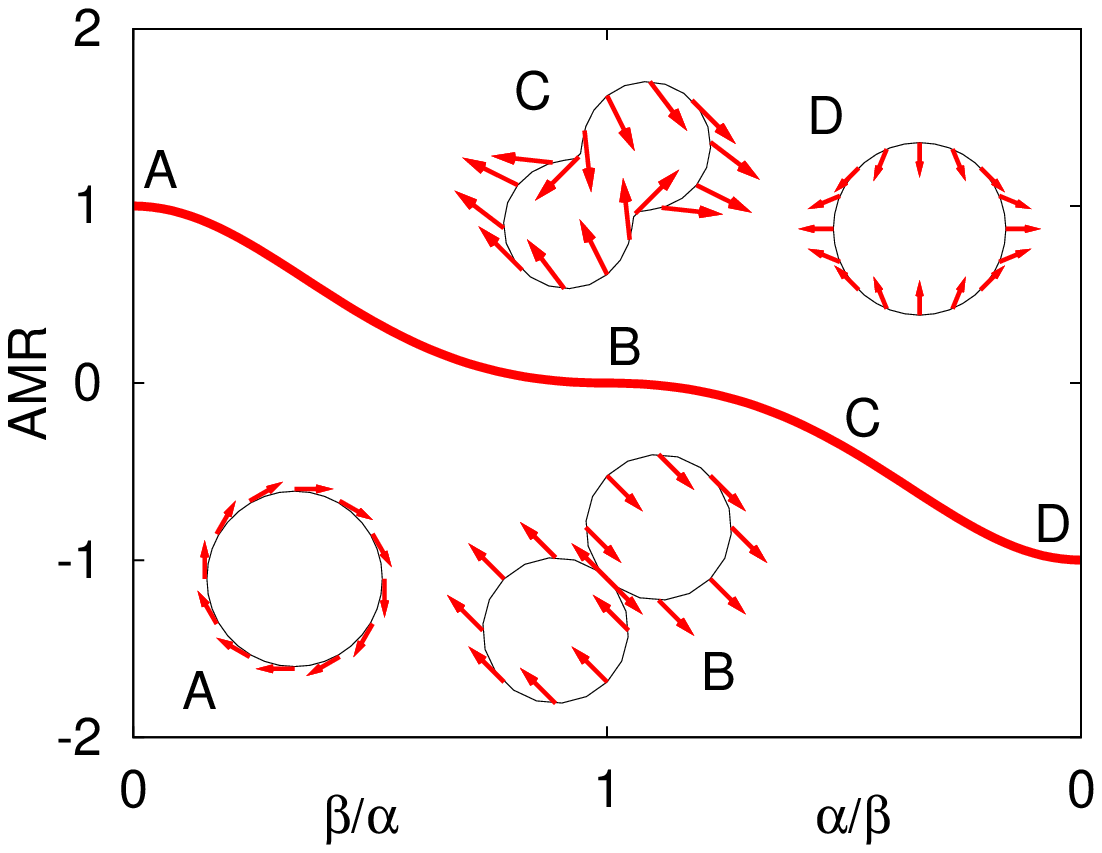} &
%\hskip-1.2cm\includegraphics[height=3.7cm]{fig2b.eps} \\
\hskip-1.2cm\includegraphics[height=3.7cm]{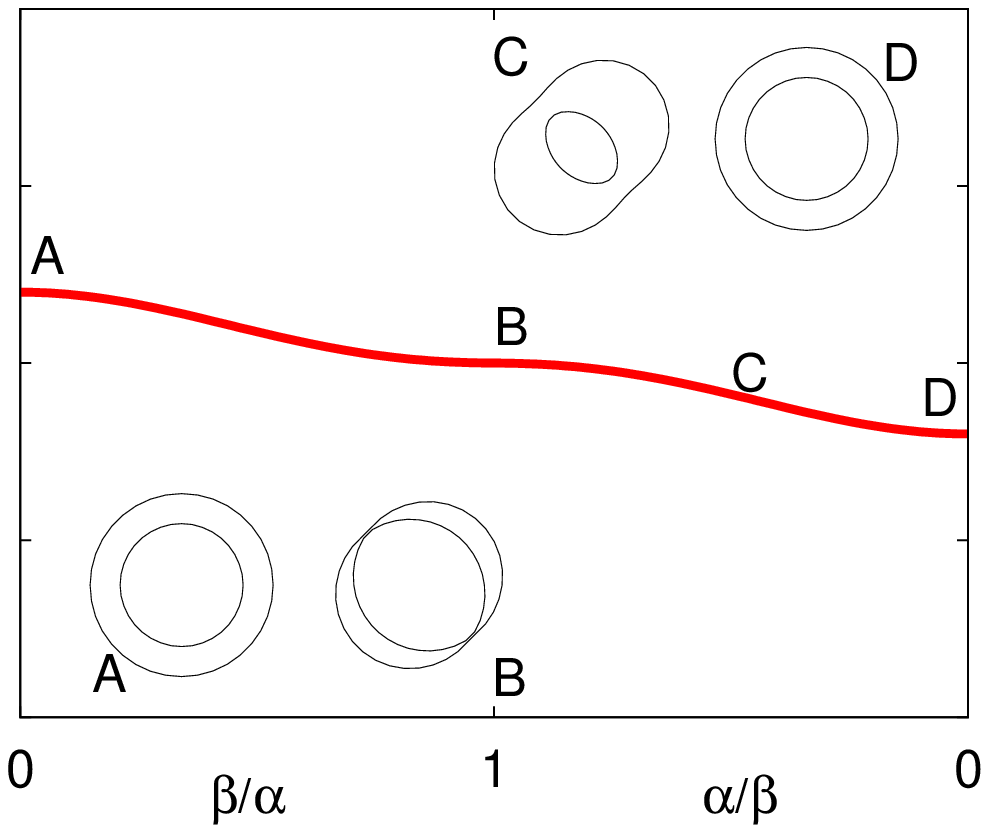} \\
(a) & (b)
\end{tabular}
\caption{AMR for pure magnetic potential impurity as a function of the ratio
  $\alpha/\beta$. Pure Rashba (Dresselhaus) interaction corresponds to the
  left (right) edge. (a) Single band case, $E_F=0$ (very low electron
  concentration). Insets show the spin textures for several
chosen values of $\alpha/\beta$. (b) Two band case with $E_F/\alpha^2>0$
fixed. Fermi lines are shown schematically,
spin textures of the majority band are qualitatively similar
to the single band case. In the limit $E_F\to\infty$, the AMR
vanishes for any value of $\alpha/\beta$.}
%\label{fig2}
\label{fig-05}
\end{figure}

The smooth transition of the AMR from the pure Rashba to pure Dresselhaus
model described by Eq.~(\ref{eq-31}) is shown in Fig.~\ref{fig-05} for $E_F=0$
and for intermediate $E_F$ corresponding to both majority
and minority Rashba-Dresselhaus bands occupied. We point out that for
$\alpha\neq\beta$ the AMR originates from not only the anisotropic
spin-texture on the Fermi contours but also, unlike the pure Rashba or pure
Dresselhaus models, from anisotropic group velocities. In the special case of
$|\alpha|=|\beta|$, these two sources of anisotropy disappear and AMR vanishes
for any {\em short-range} electro-magnetic potential.

The relative displacement along the diagonal direction of the two circular
Fermi contours is nevertheless a significant remaining imprint of the SOI in
the band structure of the $|\alpha|=|\beta|$ model. The AMR can reappear if
$w(i,\mathbf{k};i',\mathbf{k'})$ picks up a dependence on ${\mathbf k}$ and
${\mathbf k}'$ due to other than the spin-texture effect. As pointed out in
Sec.~\ref{sec-II}, a long-range (electro-)magnetic impurity potential combined
with the two displaced Fermi circles would yield wave vector dependent
$w(i,\mathbf{k};i',\mathbf{k'})$ and a non-zero AMR even 
for $|\alpha|=|\beta|$.

\section{Discussion}
\label{sec-IV}

Calculations in the previous sections show the following trends in the AMR:
(i) For the Rashba-Dresselhaus model with a short-range magnetic impurity
potential, the AMR is large (100\%) when the minority band is depleted and
when the SOI is of a pure Rashba type ($\beta=0$) or pure Dresselhaus type
($\alpha=0$). (ii) The AMR vanishes when $|\alpha|=|\beta|$ or for an
arbitrary $\alpha$ and $\beta$ when the majority and minority bands become
nearly degenerate. (iii) For impurities containing a combined electro-magnetic
potential, the AMR has the same sign as for the pure magnetic impurity
potential, is maximized when the two components have equal strength, and
remains large (200\%) even in the limit of nearly degenerate
Rashba-Dresselhaus bands. (iv) We have also noted (in agreement with
Ref.~\onlinecite{Rushforth:2007_a}) that in the higher-spin Kohn-Luttinger
model, the AMR is expected to have opposite signs for
pure magnetic potential and for electro-magnetic potential with comparable
strength of the two components. We will now discuss implications of
observations (i-iii) and inspect the applicability of our linear-response
quasiclassical theory for 2D systems with realistic material parameters.

Two-dimensional electron systems with Rashba and Dresselhaus SOI have been
studied in n-type InAs and GaAs quantum
wells\cite{Luo:1990_a,Vasilyev:2003_a,Ganichev:2004_a,Giglberger:2007_a} with
mobilities $\mu$ up to $3\times 10^5$~cm$^2$/Vs and $3.5\times
10^6$~cm$^2$/Vs, and magnitudes of the SOI of the order of $\sim 10^{-11}$~eVm
and $\sim 10^{-12}$~eVm, respectively. The ratio $|\alpha/\beta|$ is ranging
between approximately 1.5 to 8 for these two-dimensional systems with electron
densities of the order of $\sim 10^{11}-10^{12}$~cm$^{-2}$.

The semiclassical Boltzmann theory is applicable when the following two
conditions are satisfied. First, the particle's de~Broglie wavelength must be
smaller than the mean free path.  At low temperatures (as compared to the
Fermi temperature) the condition implies that
\begin{equation}
n_e>\frac{m}{\hbar\tau}\;,
\label{cond1}
\end{equation} 
where $\tau=m\mu/e$. For the above InAs and GaAs two-dimensional
systems\cite{Giglberger:2007_a} 
$m/(\hbar\tau)$ is of the order of $10^{10}\,\mathrm{cm}^{-2}$ and 
$10^9\,\mathrm{cm}^{-2}$, respectively, so the 
inequality (\ref{cond1}) can be safely met.

The second condition requires that the smearing of the spin-split bands due to
disorder is smaller than the spin splitting energy
$E_{+,\mathbf{k}}-E_{-,\mathbf{k}}$. Since the AMR we study is due to the SOI in
the band structure (rather than in the scatterers) 
it remains non-zero only in the strong SOI/weak disorder
regime.  As a consequence, the concentration must also fulfill the following
inequality
\begin{equation}\label{eq-33}
%\label{second}
n_e > \hbar^2/8\pi\kappa_\theta^2\tau^2.
\end{equation}
Assuming a pure Rashba system (i. e. $\kappa_\theta\equiv\alpha$),
the right-hand side in (\ref{eq-33}) is
of the order of $10^9\,\mathrm{cm}^{-2}$ for both InAs and
GaAs, respectively, so the condition is again satisfied for typical electron
densities. Introducing magnetic impurities will certainly decrease the
mobility of the two-dimensional systems, nevertheless, conditions
(\ref{cond1}) and (\ref{eq-33}) might remain satisfied for feasible electron
densities.  We also note that the inequality (\ref{eq-33}) can be reformulated
in terms of the mean free path $l_{\tau}$ and spin precession length,
which can be roughly estimated as $\lambda_s \sim \hbar^2/(m\alpha)$.
Namely, $l_{\tau}$ must be 
larger than $\lambda_s$ so that an electron randomizes its spin orientation
due to the spin-orbit precession between two subsequent scattering events.
This restriction corresponds to the approximation which neglects the
off-diagonal elements of the non-equilibrium distribution function in the spin
space.

Having established parameter range of the validity of the Boltzmann approach
we can now return to points (i-iii) from the beginning of this section and
comment on the expected AMRs for realistic material parameters. Since for
short range impurities the AMR is weak when $|\alpha|\approx|\beta|$ let us
assume pure Rashba model only. By a direct inspection of the results in
Tabs.~\ref{tab-02},\ref{tab-03} we find that the ratio between the isotropic
and anisotropic part, $\sigma_0$ and $\sigma_1$, of the conductivity tensor
depends on the SOI strength and electron density and can be estimated as
\begin{equation}%\label{relation}
\label{eq-34}
\frac{\sigma_1}{\sigma_0}\sim
\frac{1}{\pi n_e}\left(\frac{m\alpha}{\hbar^2}\right)^2\,.
\end{equation}
For usual electron densities $\sim 10^{11}$~cm$^{-2}$, this ratio will be of
the order of $0.01$ for a pure magnetic impurity potential, implying weak AMR
of the order of $1\%$. By depleting the minority band, the ratio
$\sigma_1/\sigma_0$ can be enhanced and the AMR can reach up to 100\% (recall
Fig.~\ref{rashba_mag_exact}). However, corresponding densities of $n_e\approx
10^{9}$~cm$^{-2}$ are relatively low compared to densities of typical
experimental two-dimensional electron systems and also we then move towards
the edge of the validity of the Boltzmann theory.

The AMR will be further reduced by the presence of another impurities than the
(electro-)magnetic ones. In terms of resistivities, this follows from the
Matthiessen's rule stating that the total resistivity is
a sum of resistivities due to the particular scattering
mechanisms.\cite{Dugdale:1967_a}  Since
scattering from pure non-magnetic impurities yields zero contribution to the
type of AMR discussed in this paper the overall relative magnetic anisotropy
of the resistivity is suppressed by their presence.

On the other hand, for impurities containing a combined electro-magnetic
potentials which add up coherently during the scattering, the AMR is expected
to be largely enhanced even in the high density regime. The strongest AMR is
predicted for similar strength of the magnetic and electric parts of the
scattering potential. This applies, e.g., to Mn in GaAs which acts both as a
charged dopant and a localized magnetic impurity, and the AMRs in GaAs:Mn can
reach $\sim 10\%$.\cite{Jungwirth:2003_b}  In the present paper, we however
wish to limit our investigation of models beyond the Rashba-Dresselhaus one to
the qualitative discussion of Sec.~\ref{sec-IIE} complemented by the exact
Boltzmann equation AMR given in Appendix~\ref{app-D}. We refer the reader to
Refs.~\onlinecite{Vyborny:2009_a,Rushforth:2007_a,Rushforth:2007_b} for a 
more quantitative discussion of AMR in (Ga,Mn)As and finally remark that
the realization of large
AMRs in Rashba-Dresselhaus systems with electro-magnetic impurities will
require doping with magnetic donors.

%\section{Summary}
%\label{sec-V}

\section*{Acknowledgements}

The work was funded through Pr\ae mium Academi\ae{} and contracts number
AV0Z10100521, LC510, KAN400100652, FON/06/E002 of \hbox{GA \v CR},
and KJB100100802 of \hbox{GA AV} of the Czech republic, by
the NAMASTE (FP7 grant No.~214499) and SemiSpinNet projects 
(FP7 grant No.~215368), by DARPA, and by DFG via SFB 689. One of us (KV) gratefully
acknowledges inspiring discussions with Vladimir I. Fal'ko
on the exact solution of the Boltzmann equation,
hospitality of Roland Winkler at the ANL, and helpful assistance
with programming of Luk\'a\v s Kripner in the adventurous period of work on
Appendix~\ref{app-D}.

\begin{appendix}

\section{Boltzmann equation for Rashba-Dresselhaus Hamiltonian}
\label{app-B}

In order to determine the non-equilibrium distribution function, we insert the
{\em ansatz}~(\ref{eq-08a}) into the Boltzmann
equation~(\ref{eq-05}) and obtain a set of four linear equations for
parameters $a^{\hat{e}_M}_{x,y}$, $b^{\hat{e}_M}_{x,y}$, one for each direction of electric field
($x,y$) and each magnetization direction $\hat{e}_M$.
For the scatterers magnetized along $\hat{x}$($\hat{y}$)-axis
and $\mathbf{E}=(E_x,0)$, we get
\begin{eqnarray}%\label{a1}
\label{eq-22a} &&
a_x=\mp\frac{\beta^2-\alpha^2}{2}\pm
\frac{1}{2}\left[a_x\frac{\beta^2-\alpha^2}{|\alpha^2-\beta^2|} \right. \\
\nonumber &&
+\left.  b_x\frac{\beta^2-\alpha^2}{2\alpha\beta}
\left(1-\frac{\alpha^2+\beta^2}{|\alpha^2 - \beta^2|}\right)
\right],\\
%\label{b1}
\label{eq-22b} &&
b_x=\mp\frac{(\alpha^2+\beta^2)^2}{4\alpha\beta}
\left(\frac{\mid \alpha^2 - \beta^2 \mid }{\alpha^2 + \beta^2}-1 \right)\\
\nonumber &&
\pm\frac{1}{2}\left[ a_x \frac{\alpha^2+\beta^2}{2\alpha\beta}
\left(1- \frac{\alpha^2+\beta^2}{\mid \alpha^2-\beta^2 \mid}\right)
+b_x\frac{\alpha^2+\beta^2}{\mid \alpha^2-\beta^2 \mid}
\right].
\end{eqnarray}

The choice $\mathbf{E}=(0,E_y)$ leads to
\begin{eqnarray}%\label{a2}
\label{eq-23a}&&
a_y=\pm\frac{\alpha^4-\beta^4}{4\alpha\beta}
\left(\frac{\mid \alpha^2 - \beta^2 \mid }{\alpha^2 + \beta^2}-1 \right)\\
\nonumber &&
\pm\frac{1}{2}\left[ a_y \frac{\beta^2-\alpha^2}{\mid \alpha^2-\beta^2 \mid}+
b_y\frac{\beta^2-\alpha^2}{2\alpha\beta}
\left(1-\frac{\alpha^2+\beta^2}{\mid \alpha^2 - \beta^2 \mid}\right)
\right],\\
%\label{b2}
\label{eq-23b}&&
b_y=\mp\frac{\alpha^2+\beta^2}{2}\pm
\frac{1}{2}\left[b_y\frac{\alpha^2+\beta^2}{|\alpha^2-\beta^2|} \right. \\
\nonumber &&
+\left.  a_y\frac{\alpha^2+\beta^2}{2\alpha\beta}
\left(1-\frac{\alpha^2+\beta^2}{|\alpha^2 - \beta^2|}\right)
\right].
\end{eqnarray}
Here, we skip the $\hat{e}_M$ superscript for brevity and relate 
the upper and lower signs to the magnetization $\mathbf{M}$ along
$\hat{x}$ and $\hat{y}$ axes respectively.

For scatterers magnetized along the $[110]$ axis we have
\begin{eqnarray}
\nonumber
&& a_x\frac{\alpha^2+\beta^2-|\alpha^2-\beta^2|}{4\alpha\beta}+\frac{b_x}{2}
+a_x=\\
%\label{axy}
\label{eq-24a}&&
\alpha\beta
\left[1+\frac{(\alpha^2+\beta^2)^2}{4\alpha^2\beta^2}\left(
\frac{|\alpha^2-\beta^2|}{\alpha^2+\beta^2}-1\right)\right],   \\
%\label{bxy}
\label{eq-24b}
&& b_x\frac{\alpha^2+\beta^2-\mid\alpha^2-\beta^2\mid}{4\alpha\beta}
+b_x + \frac{a_x}{2}=
\frac{| \alpha^2-\beta^2 |}{2},
\end{eqnarray}
while equations for $a_{y}$ and $b_{y}$ can be obtained from
Eqs.~(\ref{eq-24a},\ref{eq-24b}) by the substitution $a_{x}\rightarrow b_y$,
$b_{x}\rightarrow a_y$.

\begin{table*}
\begin{tabular}{cccc}
Magnetization & $\alpha\ge \beta$ && $\beta\ge \alpha$  \\
direction &&&\\
{[100]} &
$
\left(
\begin{array}{cc}
 -\frac{2 e^2 m \alpha ^2 (\alpha^2 - \beta^2 ) \tau }{\hbar^4 \pi  \left(3 \alpha ^2 + \beta ^2\right)} &
-\frac{2 e^2 m \alpha  (\alpha^2 - \beta^2 ) \beta   \tau }{\hbar^4 \pi  \left(3 \alpha ^2 + \beta ^2\right)} \\
-\frac{2 e^2 m \alpha  (\alpha^2 - \beta^2 ) \beta   \tau }{\hbar^4 \pi  \left(3 \alpha ^2 + \beta ^2\right)} &
-\frac{2 e^2 m \beta ^2 (\alpha^2 -\beta^2) \tau }{\hbar^4 \pi  \left(3 \alpha ^2 + \beta ^2\right)}
\end{array}
\right)
$
&&
$
\left(
\begin{array}{cc}
\frac{2 e^2 m \alpha ^2 (\alpha^2 - \beta^2 )  \tau }{\hbar^4 \pi  \left(\alpha ^2 + 3 \beta ^2\right)} &
\frac{2 e^2 m \alpha  (\alpha^2 - \beta^2 ) \beta   \tau }{\hbar^4 \pi  \left(\alpha ^2 + 3 \beta ^2\right)} \\
\frac{2 e^2 m \alpha  (\alpha^2 - \beta^2 ) \beta   \tau }{\hbar^4 \pi  \left(\alpha ^2 + 3 \beta ^2\right)}&
\frac{2 e^2 m (\alpha^2 - \beta^2 ) \beta ^2  \tau }{\hbar^4 \pi  \left(\alpha ^2 + 3 \beta ^2\right)}
\end{array}
\right)
$
\\[8mm]
{[010]} &
$
\left(
\begin{array}{cc}
-\frac{2 e^2 m \beta ^2 \left(\alpha ^2- \beta ^2 \right) \tau }{\hbar^4 \pi  \left(3 \alpha ^2 + \beta ^2\right)} &
-\frac{2 e^2 m \alpha  (\alpha^2 - \beta^2 ) \beta \tau }{\hbar^4 \pi  \left(3 \alpha ^2 + \beta ^2\right)} \\
-\frac{2 e^2 m \alpha  (\alpha^2 - \beta^2 ) \beta  \tau }{\hbar^4 \pi  \left(3 \alpha ^2 + \beta ^2\right)} &
-\frac{2 e^2 m \alpha ^2 (\alpha^2 - \beta^2 ) \tau }{\hbar^4 \pi  \left(3 \alpha ^2 + \beta ^2\right)}
\end{array}
\right)
$
&&
$\left(
\begin{array}{cc}
\frac{2 e^2 m (\alpha^2 - \beta^2 ) \beta ^2  \tau }{\hbar^4 \pi  \left(\alpha ^2 + 3 \beta ^2\right)} &
\frac{2 e^2 m \alpha  (\alpha^2 - \beta^2 ) \beta  \tau }{\hbar^4 \pi  \left(\alpha ^2 + 3 \beta ^2\right)} \\
\frac{2 e^2 m \alpha  (\alpha^2 - \beta^2 ) \beta   \tau }{\hbar^4 \pi  \left(\alpha ^2 + 3 \beta ^2\right)} &
\frac{2 e^2 m \alpha ^2 (\alpha^2 - \beta^2 )  \tau }{\hbar^4 \pi  \left(\alpha ^2 + 3 \beta ^2\right)}
\end{array}
\right)$
\\[8mm]
{[110]} &
$\left(
\begin{array}{cc}
-\frac{e^2 m (\alpha - \beta ) (\alpha + \beta )^2 \tau }{\hbar^4 \pi  (3 \alpha + \beta )} &
-\frac{e^2 m (\alpha - \beta ) (\alpha + \beta )^2 \tau }{\hbar^4 \pi  (3 \alpha + \beta )} \\
-\frac{e^2 m (\alpha - \beta ) (\alpha + \beta )^2 \tau }{\hbar^4 \pi  (3 \alpha + \beta )} &
-\frac{e^2 m (\alpha - \beta ) (\alpha + \beta )^2 \tau }{\hbar^4 \pi  (3 \alpha + \beta )}
\end{array}
\right)$
&&
$\left(
\begin{array}{cc}
\frac{e^2 m (\alpha - \beta ) (\alpha + \beta )^2 \tau }{\hbar^4 \pi  (\alpha + 3 \beta )} &
\frac{e^2 m (\alpha - \beta ) (\alpha + \beta )^2 \tau }{\hbar^4 \pi  (\alpha + 3 \beta )} \\
\frac{e^2 m (\alpha - \beta ) (\alpha + \beta )^2 \tau }{\hbar^4 \pi  (\alpha + 3 \beta )} &
\frac{e^2 m (\alpha - \beta ) (\alpha + \beta )^2 \tau }{\hbar^4 \pi  (\alpha + 3 \beta )}
\end{array}
\right)$
\\[8mm]
{[1$\bar{1}$0]} &
$\left(
\begin{array}{cc}
-\frac{e^2 m (\alpha - \beta ) (\alpha + \beta )^2 \tau }{\hbar^4 \pi  (3 \alpha + \beta )} &
\frac{e^2 m (\alpha - \beta ) (\alpha + \beta )^2 \tau }{\hbar^4 \pi  (3 \alpha + \beta )} \\
\frac{e^2 m (\alpha - \beta ) (\alpha + \beta )^2 \tau }{\hbar^4 \pi  (3 \alpha + \beta )} &
-\frac{e^2 m (\alpha - \beta ) (\alpha + \beta )^2 \tau }{\hbar^4 \pi  (3 \alpha + \beta )}
\end{array}
\right)$
&&
$\left(
\begin{array}{cc}
\frac{e^2 m (\alpha - \beta ) (\alpha + \beta )^2 \tau }{\hbar^4 \pi  (\alpha + 3 \beta )} &
-\frac{e^2 m (\alpha - \beta ) (\alpha + \beta )^2 \tau }{\hbar^4 \pi  (\alpha + 3 \beta )} \\
-\frac{e^2 m (\alpha - \beta ) (\alpha + \beta )^2 \tau }{\hbar^4 \pi  (\alpha + 3 \beta )} &
\frac{e^2 m (\alpha - \beta ) (\alpha + \beta )^2 \tau }{\hbar^4 \pi  (\alpha + 3 \beta )}
\end{array}
\right)$
%{[001]} &
%$\left(
%\begin{array}{cc}
%-\frac{2 e^2 m \left(3 \alpha ^4 - 4 \beta ^2 \alpha ^2 + \beta ^4\right) \tau }{\hbar^4 \pi  \left(9 \alpha ^2 - \beta ^2\right)} &
%\frac{4 e^2 m \alpha \beta  (\alpha^2 - \beta^2 ) \tau }{\hbar^4 \pi  \left(\beta ^2 - 9 \alpha ^2\right)} \\
%\frac{4 e^2 m \alpha  \beta  (\alpha^2 - \beta^2 ) \tau }{\hbar^4 \pi  \left(\beta ^2 - 9 \alpha ^2\right)} &
%-\frac{2 e^2 m \left(3 \alpha ^4 - 4 \beta ^2 \alpha ^2 + \beta ^4\right) \tau }{\hbar^4 \pi  \left(9 \alpha ^2 - \beta ^2\right)}
%\end{array}
%\right)$
%&&
%$\left(
%\begin{array}{cc}
%\frac{2 e^2 m \left(\alpha ^4 - 4 \beta ^2 \alpha ^2 + 3 \beta ^4\right) \tau }{\hbar^4 \pi  \left(\alpha ^2 - 9 \beta ^2\right)} &
%-\frac{4 e^2 m \alpha  \beta  (\alpha^2 - \beta^2 ) \tau }{\hbar^4 \pi  \left(\alpha ^2 - 9 \beta ^2\right)} \\
% -\frac{4 e^2 m \alpha  \beta  (\alpha^2 - \beta^2 ) \tau }{\hbar^4 \pi  \left(\alpha ^2 - 9 \beta ^2\right)} &
%\frac{2 e^2 m \left(\alpha ^4 - 4 \beta ^2 \alpha ^2 + 3 \beta ^4\right) \tau }{\hbar^4 \pi  \left(\alpha ^2 - 9 \beta ^2\right)}
%\end{array}
%\right)$
\end{tabular}
\caption{Anisotropic part $\hat{\sigma_1}$ of the total conductivity tensor
  $\hat{\sigma}=\mathds{1}\sigma_0+\hat{\sigma_1}$ for 
  arbitrary $\alpha$ and $\beta$.}
\label{tab-03}
\end{table*}

\subsection{Impurity magnetization along the [100]-axis. }

Here, we assume that impurities are magnetized along
the $x$-axis, i. e. the scattering potential
is proportional to $\sigma_x$.

If $\alpha > \beta$ then solution of Eqs.~(\ref{eq-22a},\ref{eq-22b})
and Eqs.~(\ref{eq-23a},\ref{eq-23b}) with upper sign reads
\begin{equation}\label{eq-26aA}\nonumber
a_x=\frac{\alpha ^4 - \beta ^4}{3 \alpha ^2 + \beta ^2},\quad
b_x=\frac{8 \alpha ^3 \beta }{3 \alpha ^2 + \beta ^2} - 2 \alpha  \beta ,
\end{equation}
and
\begin{equation}\label{eq-26bA}\nonumber
a_y=2 \alpha  \beta - \frac{8 \alpha ^3 \beta }{3 \alpha ^2 + \beta ^2},\quad
b_y=-\frac{24 \alpha ^4}{3 \alpha ^2 + \beta ^2} + 7 \alpha ^2 - \beta ^2\,.
\end{equation}
In the opposite case $\beta > \alpha$ the coefficients are
\begin{eqnarray}\label{eq-26aB}\nonumber
a_x=-\frac{\alpha^4 - 4 \beta^2 \alpha^2+3\beta^4}{\alpha^2+3\beta^2}, &&
b_x=\frac{2 \alpha  (\alpha^2 - \beta^2 ) \beta  }{\alpha ^2 +
      3 \beta ^2},\\
\label{eq-26bB}\nonumber
a_y=-\frac{2 \alpha  (\alpha^2 - \beta^2 ) \beta  }{\alpha ^2 +
      3 \beta ^2},&&
b_y=\frac{\beta ^4 - \alpha ^4}{\alpha ^2 + 3 \beta ^2}.
\end{eqnarray}

\subsection{Impurity magnetization along the [010]-axis.}

The scattering potential is proportional to $\sigma_y$ in this case.
If $\alpha > \beta$ then the solution of Eqs.~(\ref{eq-22a},\ref{eq-22b}) and
Eqs.~(\ref{eq-23a},\ref{eq-23b}) with lower signs is given by
\begin{eqnarray}\label{eq-27aA}\nonumber
a_x=-\frac{24 \alpha ^4}{3 \alpha ^2 + \beta^2} + 7 \alpha^2 - \beta^2, &&
b_x=2 \alpha  \beta - \frac{8 \alpha ^3 \beta }{3 \alpha ^2 + \beta ^2},\\
\label{eq-27bA}\nonumber
a_y=\frac{8 \alpha ^3 \beta }{3 \alpha ^2 + \beta ^2} - 2 \alpha\beta , &&
b_y=\frac{\alpha ^4 - \beta ^4}{3 \alpha ^2 + \beta ^2},
\end{eqnarray}
while in the opposite case ($\beta > \alpha$), the coefficients read
\begin{eqnarray}\label{eq-27aB}\nonumber
a_x=\frac{\beta ^4 - \alpha ^4}{\alpha ^2 + 3 \beta ^2}, &&
b_x=-\frac{2 \alpha  (\alpha^2 - \beta^2 ) \beta  }{\alpha ^2 +
      3 \beta ^2},\\
\label{eq-27bB}\nonumber
a_y=\frac{2 \alpha  (\alpha^2 - \beta^2 ) \beta  }{\alpha ^2 +
      3 \beta ^2},&&
b_y=-\frac{\alpha ^4 - 4 \beta ^2 \alpha^2 + 3 \beta^4}{\alpha^2 + 3 \beta^2}.
\end{eqnarray}

\subsection{Impurity magnetization along the [110]-axis.}

Here, the scattering potential is proportional to
$\frac{1}{2}(\sigma_x + \sigma_y)$.
If $\alpha > \beta$ then the coefficients $a_{x,y}$ and $b_{x,y}$ read
\begin{eqnarray}\label{eq-28aA}\nonumber
a_x=-\frac{\alpha^3 - 3\beta\alpha^2 + \beta^2\alpha+\beta^3}{3\alpha + \beta},
\quad
b_x=\frac{2 \alpha ^2 (\alpha - \beta )}{3 \alpha + \beta }
\\
\label{eq-28bA}\nonumber
a_y=\frac{2 \alpha ^2 (\alpha - \beta )}{3 \alpha + \beta },
\quad
b_y=-\frac{\alpha^3 - 3\beta\alpha^2 + \beta^2\alpha +\beta^3}{3\alpha +\beta}.
\end{eqnarray}
In the opposite case $\beta > \alpha$ we have
\begin{eqnarray}\label{eq-28aB}\nonumber
a_x=-\frac{\alpha ^3 + \beta\alpha ^2 - 3 \beta ^2 \alpha + \beta ^3}{\alpha +
      3 \beta },\quad
b_x=\frac{2 \beta ^2 (\beta - \alpha )}{\alpha + 3 \beta },
\\
\label{eq-28bB}\nonumber
a_y=\frac{2 \beta ^2 (\beta - \alpha )}{\alpha + 3 \beta },\quad
b_y=-\frac{\alpha ^3 + \beta  \alpha^2-3\beta^2\alpha+\beta^3}{\alpha+3\beta}.
\end{eqnarray}
The case when the impurities are magnetized along the [1$\bar{1}$0]-axis
can be treated in the same way.

To write down the conductivity for arbitrary $\alpha$
and $\beta$ it is convenient to define 
\begin{equation}%\label{Drude}
\label{eq-30}
\sigma_0=\frac{e^2 \tau }{m}n_e \,,
\end{equation}
where the electron concentration at $E_F\geq 0$ can be exactly expressed as
\begin{equation}\label{eq-17}
%\label{conc}
n_e=\frac{m E_F}{\pi \hbar^2} + \left(\frac{m}{\hbar^2}\right)^2
\frac{\alpha^2+\beta^2}{\pi}.
\end{equation}
Thus defined $\sigma_0$ becomes identical with the Drude
formula when $\alpha=\beta=0$. 
In fact, $\sigma_0$ times unity $2\times 2$ matrix
describes the conductivity of a 2DEG due to the non-magnetic short-range
scatterers, see
Ref.~\onlinecite{Trushin:2006_a}.  In the presence of magnetized scatterers
the conductivity acquires an additional term $\hat{\sigma_1}$ which is summarized in
Tab.~\ref{tab-03}. To obtain the total conductivity tensor one has to sum up
both these terms, i.e. $\hat{\sigma}=\mathds{1}\sigma_0+\hat{\sigma_1}$. 
Conductivity tensors
under special conditions in Tab.~\ref{tab-02} can be recovered by a proper
choice of $\alpha,\beta$ in Tab.~\ref{tab-03}.  Table~\ref{tab-03} thus
summarizes the main computational results of this paper.  They describe an
additional term in the electrical conductivity of a 2DEG confined in a
[001]-grown III-V semiconductor heterostructure due to the magnetized elastic
scatterers.

\section{Boltzmann equation for Kohn-Luttinger Hamiltonian}
\label{app-D}

Non-trivial exact analytical solutions to the Boltzmann equation~(\ref{eq-05})
exist also for some models in $d=3$ dimensions. The one described in
Section~\ref{sec-IIE} constitutes one such example and we outline here the
main steps needed to calculate the AMR in this model and thus confirm the
appropriateness of the sketch on Fig.~\ref{fig-kl}(b).

Physically, the model concerns carriers of the two heavy-hole $\Gamma_8$ bands
(HH bands) scattered off magnetic impurities. The band structure can formally
be viewed as the $(\gamma_1-2\gamma_2)/(\gamma_1+2\gamma_2)\to 0$ limit
(negligible light-hole density of states) of the
Hamiltonian~(\ref{eq-01d}) with $j_{x,y,z}=3s_{x,y,z}$ while scatterers
uniformly polarized along $z$-direction are modeled by $V\propto s_z$ in
terms of Eq.~(\ref{w}). Explicit expressions for the spin matrices $s_{x,y,z}$
can be found e.g. in the Appendix of 
Ref.~\onlinecite{Abolfath:2001_a}.  Without going into
details, we remark that this model could be used to describe the AMR in
(metallic) $p$-type III-V or II-VI semiconductors with dilute Mn impurities if
their charge is either zero or strongly 
screened;\cite{Abolfath:2001_a,Dietl:2001_b,Rushforth:2007_a}
Mn atom $d$-states hybridize with the host valence band and
create\cite{Schrieffer:1966_a} a $\vec{k}$-independent\cite{Jungwirth:2006_a}
impurity
potential $V\propto \hat{e}_M\cdot \mathbf{s}$. 
Relevant values of the proportionality constant 
and host material band structure parameters
$\gamma_1,\gamma_2$ can be found in Ref.~\onlinecite{Dietl:2001_b}.
We also stress that we will be treating a
model where the densities of states of the two involved (HH) bands are equal 
($h\to 0$) and this is of course (again) only an approximation to realistic
systems. 

Non-equilibrium distributions due to applied electric field turn out to be the
same for both HH bands in such a model, and we are required to solve
three decoupled integral equations
\begin{equation}\begin{split}
\sqrt{4\pi/3} \tau Y_1^0(\Omega)&=\bar{w}(\Omega)c(\Omega)-\int d\Omega'\,
w(\Omega,\Omega')c(\Omega')
\\ 
-\sqrt{8\pi/3} \tau Y_1^1(\Omega)&=\bar{w}(\Omega)p(\Omega)-\int d\Omega'\,
w(\Omega,\Omega')p(\Omega')
\\ 
\sqrt{8\pi/3} \tau Y_1^{-1}(\Omega)&=\bar{w}(\Omega)q(\Omega)-\int d\Omega'\,
w(\Omega,\Omega')q(\Omega')
\end{split}\label{eq-36} 
\end{equation}
where $\tau$ is defined by the 3D analogy of Eq.~(\ref{tau}), 
%$1/\tau=2\pi J_{pd}n/\hbar$ 
$\Omega$ denotes a compound variable $\varphi,\vartheta$ parameterizing
the unit sphere, so that $\int d\Omega = \int_{0}^{\pi}\sin\vartheta\
d\vartheta \int_0^{2\pi} d\varphi$, 
\begin{widetext}
\begin{eqnarray}  \nonumber
  w(\Omega,\Omega') &= &
  - \frac{2\pi}{15}\left( Y_2^{-1}{Y_2^1}'+Y_2^1{Y_2^{-1}}' \right)
 +\frac{10\pi}{9}Y_0^0{Y_0^0}' 
 +\frac{8\pi}{9\sqrt{5}} \bigg( Y_0^0{Y_2^0}'+  Y_2^0{Y_0^0}' \bigg)
 + \frac{2\pi}{9}Y_2^0{Y_2^0}' 
 - \frac{2\pi}{15}\left( Y_2^{-2}{Y_2^2}'+Y_2^2{Y_2^{-2}}'\right)
\\ \label{eq-37}
  \bar{w}(\Omega) &=&  \int d\Omega'\, w(\Omega,\Omega') =
    \frac{2\pi\sqrt{4\pi }}{3} 
  \left(\frac{4}{3\sqrt{5}} Y_2^0 + \frac{5}{3}Y_0^0\right)\,,
\end{eqnarray}
and $Y_l^m=Y_l^m(\Omega)$, $Y_l^{m'}=Y_l^m(\Omega')$ denote spherical harmonics
normalized to $\int d\Omega\, Y_l^{m,*}(\Omega)
Y_{l'}^{m'}(\Omega)=\delta_{ll'}\delta_{mm'}$ according to the Condon-Shortley
convention. 

Non-equilibrium distribution under the effect of $\mathbf{E}=(E_x,E_y,E_z)$ is
then 
\begin{equation}
%\begin{split}
  f(\mathbf{k})-f^0(E_{k}) = 
  e v \frac{\partial f^0(E_k)}{\partial E_k} \Bigg( \textstyle
  \frac{1}{2} \big(p(\Omega)+q(\Omega)\big)E_x - %\\
  \frac{1}{2}i \big(p(\Omega)-q(\Omega)\big) E_y + c(\Omega) E_z  \Bigg)
%\end{split}
\end{equation}
\end{widetext}
for the both bands (which are in the $h\to 0$ approximation identical),
and conductivities implied in the spirit of Eq.~(\ref{current}) are
\begin{eqnarray*}
  \sigma_{xx}&=&\int d\Omega\, \frac{1}{8}\big(p(\Omega)+q(\Omega)\big)
  \big(Y_1^{-1}(\Omega)-Y_1^{1}(\Omega)\big)\\
  \sigma_{yy}&=&\int d\Omega\, \frac{1}{8}\big(p(\Omega)-q(\Omega)\big)
  \big(Y_1^1(\Omega)+Y_1^{-1}(\Omega)\big)\\
  \sigma_{zz}&=&\int d\Omega\, \frac{\sqrt{2}}{4}c(\Omega) Y_1^0(\Omega)\,.
  \end{eqnarray*}
in units of $\sqrt{3/2\pi}\cdot e^2n_e/m$.
Using analytical solutions of Eqs.~(\ref{eq-36}) corresponding to scattering
amplitudes~(\ref{eq-37}), we find 
\begin{equation}\label{theveryverylastandlongexpectedresultGLORIAgloria}
  \mbox{AMR} = 2\frac{-6+5\sqrt{2}\arctan\sqrt{2}}{2+\sqrt{2}\arctan\sqrt{2}}
        \approx 0.45\,,
\end{equation}
according to our definition of AMR~(\ref{eq-03}),
that is resistance parallel to the magnetization is higher.  It is thus
confirmed that sketches for pure magnetic scattering in Fig.~\ref{fig-kl}(b) 
appropriately describe conductivity calculated by exactly solving the
Boltzmann equation. We also obtained the same 
result~(\ref{theveryverylastandlongexpectedresultGLORIAgloria}) 
within the Keldysh
formalism\cite{Kovalev:2009_a,Kovalev:2008_a} where conductivity $\sigma_{xx}$
turns out to be proportional to $\langle v_x\delta v_x\delta\rangle_t$ with
$v_x=\partial H_{KL}/\partial k_x$, $\delta=G^R-G^A$, $G^A=(G^R)^\dag$ and
$(G^R)^{-1}=E_F - H_{KL}-\Sigma^R$ with self-energy $\Sigma^R\propto
s_z^2$. Brackets $\langle \ldots \rangle_t$ in this Kubo-formula-type
result\cite{streda} mean
trace in the space of $4\times 4$ matrices and integration over the
$k$-space. Conductivity $\sigma_{zz}$ is analogous ($v_x$ is replaced by
$v_z$, $\Sigma^R$ stays unchanged).

We finally remark that Eqs.~(\ref{eq-36}) are completely analogous to the two
equations~(8,7) of the 2D case in Ref.~\onlinecite{Vyborny:2008_a}. Solution
of those equations was constructed in the form of a Fourier series or a
modified Fourier series as explained in the note added in proof of that
reference. In our current 3D problem defined by Eqs.~(\ref{eq-36},\ref{eq-37}),
if expanded in terms of modified spherical harmonics
$Y_l^m(\Omega)/\bar{w}(\Omega)$, the solutions $c(\Omega)$, $p(\Omega)$,
$q(\Omega)$ are found to contain only few terms [the harmonics present in the
left-hand-sides of Eqs.~(\ref{eq-36}); note the analogy to the discussion of
Sec.~\ref{sec-IIIA} and Eq.~(\ref{eq-08a}) when spherical harmonics
replace sines and cosines]. During the calculations we
have to be however cautious as the integrals of Eqs.~(\ref{eq-36}) do not
contain simple scalar products of spherical harmonics where orthogonality
relations apply.

\end{appendix}

%\bibliography{MSWEBpublications}

\begin{thebibliography}{47}
\expandafter\ifx\csname natexlab\endcsname\relax\def\natexlab#1{#1}\fi
\expandafter\ifx\csname bibnamefont\endcsname\relax
  \def\bibnamefont#1{#1}\fi
\expandafter\ifx\csname bibfnamefont\endcsname\relax
  \def\bibfnamefont#1{#1}\fi
\expandafter\ifx\csname citenamefont\endcsname\relax
  \def\citenamefont#1{#1}\fi
\expandafter\ifx\csname url\endcsname\relax
  \def\url#1{\texttt{#1}}\fi
\expandafter\ifx\csname urlprefix\endcsname\relax\def\urlprefix{URL }\fi
\providecommand{\bibinfo}[2]{#2}
\providecommand{\eprint}[2][]{\url{#2}}

\bibitem[{\citenamefont{Chien and Westgate}(1980)}]{Chien:1980_a}
\bibinfo{author}{\bibfnamefont{L.}~\bibnamefont{Chien}} \bibnamefont{and}
  \bibinfo{author}{\bibfnamefont{C.~R.} \bibnamefont{Westgate}},
  \emph{\bibinfo{title}{The Hall Effect and Its Applications}}
  (\bibinfo{publisher}{Plenum, New York}, \bibinfo{year}{1980}).

\bibitem[{\citenamefont{Sinova et~al.}(2004)\citenamefont{Sinova, Jungwirth,
  and {\v{C}erne}}}]{Sinova:2004_c}
\bibinfo{author}{\bibfnamefont{J.}~\bibnamefont{Sinova}},
  \bibinfo{author}{\bibfnamefont{T.}~\bibnamefont{Jungwirth}},
  \bibnamefont{and}
  \bibinfo{author}{\bibfnamefont{J.}~\bibnamefont{{\v{C}erne}}},
  \bibinfo{journal}{Int. J. Mod. Phys.} \textbf{\bibinfo{volume}{B 18}},
  \bibinfo{pages}{1083} (\bibinfo{year}{2004}),
  \eprint{arXiv:cond-mat/0402568}.

\bibitem[{\citenamefont{Sinitsyn}(2008)}]{Sinitsyn:2007_a}
\bibinfo{author}{\bibfnamefont{N.~A.} \bibnamefont{Sinitsyn}},
  \bibinfo{journal}{J. Phys.: Condens. Matter} \textbf{\bibinfo{volume}{20}},
  \bibinfo{pages}{023201} (\bibinfo{year}{2008}), \eprint{arXiv:0712.0183}.

\bibitem[{\citenamefont{Jungwirth
  et~al.}(2002{\natexlab{a}})\citenamefont{Jungwirth, Niu, and
  MacDonald}}]{Jungwirth:2002_a}
\bibinfo{author}{\bibfnamefont{T.}~\bibnamefont{Jungwirth}},
  \bibinfo{author}{\bibfnamefont{Q.}~\bibnamefont{Niu}}, \bibnamefont{and}
  \bibinfo{author}{\bibfnamefont{A.~H.} \bibnamefont{MacDonald}},
  \bibinfo{journal}{Phys. Rev. Lett.} \textbf{\bibinfo{volume}{88}},
  \bibinfo{pages}{207208} (\bibinfo{year}{2002}{\natexlab{a}}),
  \eprint{arXiv:cond-mat/0110484}.

\bibitem[{\citenamefont{Dietl et~al.}(2003)\citenamefont{Dietl, Matsukura,
  Ohno, Cibert, and Ferrand}}]{Dietl:2003_c}
\bibinfo{author}{\bibfnamefont{T.}~\bibnamefont{Dietl}},
  \bibinfo{author}{\bibfnamefont{F.}~\bibnamefont{Matsukura}},
  \bibinfo{author}{\bibfnamefont{H.}~\bibnamefont{Ohno}},
  \bibinfo{author}{\bibfnamefont{J.}~\bibnamefont{Cibert}}, \bibnamefont{and}
  \bibinfo{author}{\bibfnamefont{D.}~\bibnamefont{Ferrand}}, in
  \emph{\bibinfo{booktitle}{Recent Trends in Theory of Physical Phenomena in
  High Magnetic Fields}}, edited by
  \bibinfo{editor}{\bibfnamefont{I.}~\bibnamefont{Vagner}}
  (\bibinfo{publisher}{Kluwer, Dordrecht}, \bibinfo{year}{2003}), p.
  \bibinfo{pages}{197}, \eprint{arXiv:cond-mat/0306484}.

\bibitem[{\citenamefont{Ruzmetov et~al.}(2004)\citenamefont{Ruzmetov,
  Scherschligt, Baxter, Wojtowicz, Liu, Sasaki, Furdyna, Yu, and
  Walukiewicz}}]{Ruzmetov:2004_a}
\bibinfo{author}{\bibfnamefont{D.}~\bibnamefont{Ruzmetov}},
  \bibinfo{author}{\bibfnamefont{J.}~\bibnamefont{Scherschligt}},
  \bibinfo{author}{\bibfnamefont{D.~V.} \bibnamefont{Baxter}},
  \bibinfo{author}{\bibfnamefont{T.}~\bibnamefont{Wojtowicz}},
  \bibinfo{author}{\bibfnamefont{X.}~\bibnamefont{Liu}},
  \bibinfo{author}{\bibfnamefont{Y.}~\bibnamefont{Sasaki}},
  \bibinfo{author}{\bibfnamefont{J.~K.} \bibnamefont{Furdyna}},
  \bibinfo{author}{\bibfnamefont{K.~M.} \bibnamefont{Yu}}, \bibnamefont{and}
  \bibinfo{author}{\bibfnamefont{W.}~\bibnamefont{Walukiewicz}},
  \bibinfo{journal}{Phys. Rev.} \textbf{\bibinfo{volume}{B 69}},
  \bibinfo{pages}{155207} (\bibinfo{year}{2004}).

\bibitem[{\citenamefont{Chun et~al.}(2007)\citenamefont{Chun, Kim, Choi, Jeong,
  Lee, Suh, Oh, Kim, Khim, Woo et~al.}}]{Chun:2006_a}
\bibinfo{author}{\bibfnamefont{S.~H.} \bibnamefont{Chun}},
  \bibinfo{author}{\bibfnamefont{Y.~S.} \bibnamefont{Kim}},
  \bibinfo{author}{\bibfnamefont{H.~K.} \bibnamefont{Choi}},
  \bibinfo{author}{\bibfnamefont{I.~T.} \bibnamefont{Jeong}},
  \bibinfo{author}{\bibfnamefont{W.~O.} \bibnamefont{Lee}},
  \bibinfo{author}{\bibfnamefont{K.~S.} \bibnamefont{Suh}},
  \bibinfo{author}{\bibfnamefont{Y.~S.} \bibnamefont{Oh}},
  \bibinfo{author}{\bibfnamefont{K.~H.} \bibnamefont{Kim}},
  \bibinfo{author}{\bibfnamefont{Z.~G.} \bibnamefont{Khim}},
  \bibinfo{author}{\bibfnamefont{J.~C.} \bibnamefont{Woo}},
  \bibnamefont{et~al.}, \bibinfo{journal}{Phys. Rev. Lett.}
  \textbf{\bibinfo{volume}{98}}, \bibinfo{pages}{026601}
  (\bibinfo{year}{2007}), \eprint{arXiv:cond-mat/0603808}.

\bibitem[{\citenamefont{{Mih\'{a}ly} et~al.}(2008)\citenamefont{{Mih\'{a}ly},
  Csontos, {Bord\'{a}cs}, {K\'{e}zsm\'{a}rki}, Wojtowicz, Liu, {Jank\'{o}}, and
  Furdyna}}]{Mihaly:2007_a}
\bibinfo{author}{\bibfnamefont{G.}~\bibnamefont{{Mih\'{a}ly}}},
  \bibinfo{author}{\bibfnamefont{M.}~\bibnamefont{Csontos}},
  \bibinfo{author}{\bibfnamefont{S.}~\bibnamefont{{Bord\'{a}cs}}},
  \bibinfo{author}{\bibfnamefont{I.}~\bibnamefont{{K\'{e}zsm\'{a}rki}}},
  \bibinfo{author}{\bibfnamefont{T.}~\bibnamefont{Wojtowicz}},
  \bibinfo{author}{\bibfnamefont{X.}~\bibnamefont{Liu}},
  \bibinfo{author}{\bibfnamefont{B.}~\bibnamefont{{Jank\'{o}}}},
  \bibnamefont{and} \bibinfo{author}{\bibfnamefont{J.~K.}
  \bibnamefont{Furdyna}}, \bibinfo{journal}{Phys. Rev. Lett.}
  \textbf{\bibinfo{volume}{100}}, \bibinfo{pages}{107201}
  (\bibinfo{year}{2008}), \eprint{arXiv:0709.0059}.

\bibitem[{\citenamefont{Arsenault et~al.}(2008)\citenamefont{Arsenault,
  Movaghar, Desjardins, and Yelon}}]{Arsenault:2008_b}
\bibinfo{author}{\bibfnamefont{L.-F.} \bibnamefont{Arsenault}},
  \bibinfo{author}{\bibfnamefont{B.}~\bibnamefont{Movaghar}},
  \bibinfo{author}{\bibfnamefont{P.}~\bibnamefont{Desjardins}},
  \bibnamefont{and} \bibinfo{author}{\bibfnamefont{A.}~\bibnamefont{Yelon}},
  \bibinfo{journal}{Phys. Rev.} \textbf{\bibinfo{volume}{B 77}},
  \bibinfo{pages}{115211} (\bibinfo{year}{2008}), \eprint{arXiv:0801.1840}.

\bibitem[{\citenamefont{Pu et~al.}(2008)\citenamefont{Pu, Chiba, Matsukura,
  Ohno, and Shi}}]{Pu:2008_a}
\bibinfo{author}{\bibfnamefont{Y.}~\bibnamefont{Pu}},
  \bibinfo{author}{\bibfnamefont{D.}~\bibnamefont{Chiba}},
  \bibinfo{author}{\bibfnamefont{F.}~\bibnamefont{Matsukura}},
  \bibinfo{author}{\bibfnamefont{H.}~\bibnamefont{Ohno}}, \bibnamefont{and}
  \bibinfo{author}{\bibfnamefont{J.}~\bibnamefont{Shi}},
  \bibinfo{journal}{Phys. Rev. Lett.} \textbf{\bibinfo{volume}{101}},
  \bibinfo{pages}{117208} (\bibinfo{year}{2008}), \eprint{arXiv:0807.4942}.

\bibitem[{\citenamefont{Matsukura et~al.}(2002)\citenamefont{Matsukura, Ohno,
  and Dietl}}]{Matsukura:2002_a}
\bibinfo{author}{\bibfnamefont{F.}~\bibnamefont{Matsukura}},
  \bibinfo{author}{\bibfnamefont{H.}~\bibnamefont{Ohno}}, \bibnamefont{and}
  \bibinfo{author}{\bibfnamefont{T.}~\bibnamefont{Dietl}}, in
  \emph{\bibinfo{booktitle}{Handbook of Magnetic Materials}}, edited by
  \bibinfo{editor}{\bibfnamefont{K.~H.~J.} \bibnamefont{Buschow}}
  (\bibinfo{publisher}{Elsevier, Amsterdam}, \bibinfo{year}{2002}),
  vol.~\bibinfo{volume}{14}, p.~\bibinfo{pages}{1}, \eprint{From Ohno Lab
  Homepage}.

\bibitem[{\citenamefont{Jungwirth et~al.}(2006)\citenamefont{Jungwirth, Sinova,
  {Ma\v{s}ek}, {Ku\v{c}era}, and MacDonald}}]{Jungwirth:2006_a}
\bibinfo{author}{\bibfnamefont{T.}~\bibnamefont{Jungwirth}},
  \bibinfo{author}{\bibfnamefont{J.}~\bibnamefont{Sinova}},
  \bibinfo{author}{\bibfnamefont{J.}~\bibnamefont{{Ma\v{s}ek}}},
  \bibinfo{author}{\bibfnamefont{J.}~\bibnamefont{{Ku\v{c}era}}},
  \bibnamefont{and} \bibinfo{author}{\bibfnamefont{A.~H.}
  \bibnamefont{MacDonald}}, \bibinfo{journal}{Rev. Mod. Phys.}
  \textbf{\bibinfo{volume}{78}}, \bibinfo{pages}{809} (\bibinfo{year}{2006}),
  \eprint{arXiv:cond-mat/0603380}.

\bibitem[{\citenamefont{Dugaev et~al.}(2005)\citenamefont{Dugaev, Bruno,
  Taillefumier, Canals, and Lacroix}}]{Dugaev:2005_a}
\bibinfo{author}{\bibfnamefont{V.~K.} \bibnamefont{Dugaev}},
  \bibinfo{author}{\bibfnamefont{P.}~\bibnamefont{Bruno}},
  \bibinfo{author}{\bibfnamefont{M.}~\bibnamefont{Taillefumier}},
  \bibinfo{author}{\bibfnamefont{B.}~\bibnamefont{Canals}}, \bibnamefont{and}
  \bibinfo{author}{\bibfnamefont{C.}~\bibnamefont{Lacroix}},
  \bibinfo{journal}{Phys. Rev.} \textbf{\bibinfo{volume}{B 71}},
  \bibinfo{pages}{224423} (\bibinfo{year}{2005}),
  \eprint{arXiv:cond-mat/0502386}.

\bibitem[{\citenamefont{Trushin and Schliemann}(2007)}]{Trushin:2006_a}
\bibinfo{author}{\bibfnamefont{M.}~\bibnamefont{Trushin}} \bibnamefont{and}
  \bibinfo{author}{\bibfnamefont{J.}~\bibnamefont{Schliemann}},
  \bibinfo{journal}{Phys. Rev.} \textbf{\bibinfo{volume}{B 75}},
  \bibinfo{pages}{155323} (\bibinfo{year}{2007}),
  \eprint{arXiv:cond-mat/0611328}.

\bibitem[{\citenamefont{Sinitsyn et~al.}(2007)\citenamefont{Sinitsyn,
  MacDonald, Jungwirth, Dugaev, and Sinova}}]{Sinitsyn:2006_b}
\bibinfo{author}{\bibfnamefont{N.~A.} \bibnamefont{Sinitsyn}},
  \bibinfo{author}{\bibfnamefont{A.~H.} \bibnamefont{MacDonald}},
  \bibinfo{author}{\bibfnamefont{T.}~\bibnamefont{Jungwirth}},
  \bibinfo{author}{\bibfnamefont{V.~K.} \bibnamefont{Dugaev}},
  \bibnamefont{and} \bibinfo{author}{\bibfnamefont{J.}~\bibnamefont{Sinova}},
  \bibinfo{journal}{Phys. Rev.} \textbf{\bibinfo{volume}{B 75}},
  \bibinfo{pages}{045315} (\bibinfo{year}{2007}),
  \eprint{arXiv:cond-mat/0608682}.

\bibitem[{\citenamefont{ichiro Inoue et~al.}(2006)\citenamefont{ichiro Inoue,
  Kato, Ishikawa, Itoh, Bauer, and Molenkamp}}]{Inoue:2006_a}
\bibinfo{author}{\bibfnamefont{J.-I.}~\bibnamefont{Inoue}},
  \bibinfo{author}{\bibfnamefont{T.}~\bibnamefont{Kato}},
  \bibinfo{author}{\bibfnamefont{Y.}~\bibnamefont{Ishikawa}},
  \bibinfo{author}{\bibfnamefont{H.}~\bibnamefont{Itoh}},
  \bibinfo{author}{\bibfnamefont{G.~E.~W.} \bibnamefont{Bauer}},
  \bibnamefont{and} \bibinfo{author}{\bibfnamefont{L.~W.}
  \bibnamefont{Molenkamp}}, \bibinfo{journal}{Phys. Rev. Lett.}
  \textbf{\bibinfo{volume}{97}}, \bibinfo{pages}{046604}
  (\bibinfo{year}{2006}), \eprint{arXiv:cond-mat/0604108}.

\bibitem[{\citenamefont{Onoda et~al.}(2008)\citenamefont{Onoda, Sugimoto, and
  Nagaosa}}]{Onoda:2007_a}
\bibinfo{author}{\bibfnamefont{S.}~\bibnamefont{Onoda}},
  \bibinfo{author}{\bibfnamefont{N.}~\bibnamefont{Sugimoto}}, \bibnamefont{and}
  \bibinfo{author}{\bibfnamefont{N.}~\bibnamefont{Nagaosa}},
  \bibinfo{journal}{Phys. Rev.} \textbf{\bibinfo{volume}{B 77}},
  \bibinfo{pages}{165103} (\bibinfo{year}{2008}), \eprint{arXiv:0712.0210}.

\bibitem[{\citenamefont{Nunner et~al.}(2007)\citenamefont{Nunner, Sinitsyn,
  Borunda, Dugaev, Kovalev, Abanov, Timm, Jungwirth, ichiro Inoue, MacDonald
  et~al.}}]{Nunner:2007_a}
\bibinfo{author}{\bibfnamefont{T.~S.} \bibnamefont{Nunner}},
  \bibinfo{author}{\bibfnamefont{N.~A.} \bibnamefont{Sinitsyn}},
  \bibinfo{author}{\bibfnamefont{M.~F.} \bibnamefont{Borunda}},
  \bibinfo{author}{\bibfnamefont{V.~K.} \bibnamefont{Dugaev}},
  \bibinfo{author}{\bibfnamefont{A.~A.} \bibnamefont{Kovalev}},
  \bibinfo{author}{\bibfnamefont{A.}~\bibnamefont{Abanov}},
  \bibinfo{author}{\bibfnamefont{C.}~\bibnamefont{Timm}},
  \bibinfo{author}{\bibfnamefont{T.}~\bibnamefont{Jungwirth}},
  \bibinfo{author}{\bibfnamefont{J.}~\bibnamefont{ichiro Inoue}},
  \bibinfo{author}{\bibfnamefont{A.~H.} \bibnamefont{MacDonald}},
  \bibnamefont{et~al.}, \bibinfo{journal}{Phys. Rev.}
  \textbf{\bibinfo{volume}{B 76}}, \bibinfo{pages}{235312}
  (\bibinfo{year}{2007}), \eprint{arXiv:0706.0056}.

\bibitem[{\citenamefont{Nunner et~al.}(2008)\citenamefont{Nunner, {Zar\'{a}nd},
  and von Oppen}}]{Nunner:2007_b}
\bibinfo{author}{\bibfnamefont{T.~S.} \bibnamefont{Nunner}},
  \bibinfo{author}{\bibfnamefont{G.}~\bibnamefont{{Zar\'{a}nd}}},
  \bibnamefont{and} \bibinfo{author}{\bibfnamefont{F.}~\bibnamefont{von
  Oppen}}, \bibinfo{journal}{Phys. Rev. Lett.} \textbf{\bibinfo{volume}{100}},
  \bibinfo{pages}{236602} (\bibinfo{year}{2008}), \eprint{arXiv:0711.3415}.

\bibitem[{\citenamefont{Borunda et~al.}(2007)\citenamefont{Borunda, Nunner,
  Luck, Sinitsyn, Timm, Wunderlich, Jungwirth, MacDonald, and
  Sinova}}]{Borunda:2007_a}
\bibinfo{author}{\bibfnamefont{M.}~\bibnamefont{Borunda}},
  \bibinfo{author}{\bibfnamefont{T.~S.} \bibnamefont{Nunner}},
  \bibinfo{author}{\bibfnamefont{T.}~\bibnamefont{Luck}},
  \bibinfo{author}{\bibfnamefont{N.~A.} \bibnamefont{Sinitsyn}},
  \bibinfo{author}{\bibfnamefont{C.}~\bibnamefont{Timm}},
  \bibinfo{author}{\bibfnamefont{J.}~\bibnamefont{Wunderlich}},
  \bibinfo{author}{\bibfnamefont{T.}~\bibnamefont{Jungwirth}},
  \bibinfo{author}{\bibfnamefont{A.~H.} \bibnamefont{MacDonald}},
  \bibnamefont{and} \bibinfo{author}{\bibfnamefont{J.}~\bibnamefont{Sinova}},
  \bibinfo{journal}{Phys. Rev. Lett.} \textbf{\bibinfo{volume}{99}},
  \bibinfo{pages}{066604} (\bibinfo{year}{2007}),
  \eprint{arXiv:cond-mat/0702289}.

\bibitem[{\citenamefont{Kovalev et~al.}(2008)\citenamefont{Kovalev,
  {V\'{y}born\'{y}}, and Sinova}}]{Kovalev:2008_a}
\bibinfo{author}{\bibfnamefont{A.~A.} \bibnamefont{Kovalev}},
  \bibinfo{author}{\bibfnamefont{K.}~\bibnamefont{{V\'{y}born\'{y}}}},
  \bibnamefont{and} \bibinfo{author}{\bibfnamefont{J.}~\bibnamefont{Sinova}},
  \bibinfo{journal}{Phys. Rev.} \textbf{\bibinfo{volume}{B 78}},
  \bibinfo{pages}{041305} (\bibinfo{year}{2008}), \eprint{arXiv:0803.1226}.

\bibitem[{\citenamefont{Sinova and MacDonald}(2008)}]{Sinova:2008_a}
\bibinfo{author}{\bibfnamefont{J.}~\bibnamefont{Sinova}} \bibnamefont{and}
  \bibinfo{author}{\bibfnamefont{A.~H.} \bibnamefont{MacDonald}}, in
  \emph{\bibinfo{booktitle}{Spintronics edited by
  Tomasz Dietl, David D. Awschalom,  Maria Kaminska,  and Hideo Ohno}}
  (\bibinfo{publisher}{Elsevier}, \bibinfo{year}{2008}),
  vol.~\bibinfo{volume}{82} of \emph{\bibinfo{series}{Semicond. Semimet.}},
  p.~\bibinfo{pages}{45}.

\bibitem[{\citenamefont{Kovalev et~al.}(2009)\citenamefont{Kovalev,
  Tserkovnyak, Vyborny, and Sinova}}]{Kovalev:2009_a}
\bibinfo{author}{\bibfnamefont{A.~A.} \bibnamefont{Kovalev}},
  \bibinfo{author}{\bibfnamefont{Y.}~\bibnamefont{Tserkovnyak}},
  \bibinfo{author}{\bibfnamefont{K.}~\bibnamefont{Vyborny}}, \bibnamefont{and}
  \bibinfo{author}{\bibfnamefont{J.}~\bibnamefont{Sinova}},
  \bibinfo{journal}{Phys. Rev.} \textbf{\bibinfo{volume}{B 79}},
  \bibinfo{pages}{195129} (\bibinfo{year}{2009}), \eprint{arXiv:0902.2571}.

\bibitem[{\citenamefont{Smit}(1951)}]{Smit:1951_a}
\bibinfo{author}{\bibfnamefont{J.}~\bibnamefont{Smit}},
  \bibinfo{journal}{Physica} \textbf{\bibinfo{volume}{17}},
  \bibinfo{pages}{612} (\bibinfo{year}{1951}).

\bibitem[{\citenamefont{Jaoul et~al.}(1977)\citenamefont{Jaoul, Campbell, and
  Fert}}]{Jaoul:1977_a}
\bibinfo{author}{\bibfnamefont{O.}~\bibnamefont{Jaoul}},
  \bibinfo{author}{\bibfnamefont{I.~A.} \bibnamefont{Campbell}},
  \bibnamefont{and} \bibinfo{author}{\bibfnamefont{A.}~\bibnamefont{Fert}},
  \bibinfo{journal}{J. Magn. Magn. Mater.} \textbf{\bibinfo{volume}{5}},
  \bibinfo{pages}{23} (\bibinfo{year}{1977}).

\bibitem[{\citenamefont{McGuire and Potter}(1975)}]{McGuire:1975_a}
\bibinfo{author}{\bibfnamefont{T.}~\bibnamefont{McGuire}} \bibnamefont{and}
  \bibinfo{author}{\bibfnamefont{R.}~\bibnamefont{Potter}},
  \bibinfo{journal}{IEEE Trans. Magn.} \textbf{\bibinfo{volume}{11}},
  \bibinfo{pages}{1018} (\bibinfo{year}{1975}).

\bibitem[{\citenamefont{Banhart and Ebert}(1995)}]{Banhart:1995_a}
\bibinfo{author}{\bibfnamefont{J.}~\bibnamefont{Banhart}} \bibnamefont{and}
  \bibinfo{author}{\bibfnamefont{H.}~\bibnamefont{Ebert}},
  \bibinfo{journal}{Europhys. Lett.} \textbf{\bibinfo{volume}{32}},
  \bibinfo{pages}{517} (\bibinfo{year}{1995}).

\bibitem[{\citenamefont{Khmelevskyi et~al.}(2003)\citenamefont{Khmelevskyi,
  {Palot\'{a}s}, Szunyogh, and Weinberger}}]{Khmelevskyi:2003_a}
\bibinfo{author}{\bibfnamefont{S.}~\bibnamefont{Khmelevskyi}},
  \bibinfo{author}{\bibfnamefont{K.}~\bibnamefont{{Palot\'{a}s}}},
  \bibinfo{author}{\bibfnamefont{L.}~\bibnamefont{Szunyogh}}, \bibnamefont{and}
  \bibinfo{author}{\bibfnamefont{P.}~\bibnamefont{Weinberger}},
  \bibinfo{journal}{Phys. Rev.} \textbf{\bibinfo{volume}{B 68}},
  \bibinfo{pages}{012402} (\bibinfo{year}{2003}).

\bibitem[{\citenamefont{Rushforth et~al.}(2007)\citenamefont{Rushforth,
  {V\'{y}born\'{y}}, King, Edmonds, Campion, Foxon, Wunderlich, Irvine,
  {Va\v{s}ek}, {Nov\'{a}k} et~al.}}]{Rushforth:2007_a}
\bibinfo{author}{\bibfnamefont{A.~W.} \bibnamefont{Rushforth}},
  \bibinfo{author}{\bibfnamefont{K.}~\bibnamefont{{V\'{y}born\'{y}}}},
  \bibinfo{author}{\bibfnamefont{C.~S.} \bibnamefont{King}},
  \bibinfo{author}{\bibfnamefont{K.~W.} \bibnamefont{Edmonds}},
  \bibinfo{author}{\bibfnamefont{R.~P.} \bibnamefont{Campion}},
  \bibinfo{author}{\bibfnamefont{C.~T.} \bibnamefont{Foxon}},
  \bibinfo{author}{\bibfnamefont{J.}~\bibnamefont{Wunderlich}},
  \bibinfo{author}{\bibfnamefont{A.~C.} \bibnamefont{Irvine}},
  \bibinfo{author}{\bibfnamefont{P.}~\bibnamefont{{Va\v{s}ek}}},
  \bibinfo{author}{\bibfnamefont{V.}~\bibnamefont{{Nov\'{a}k}}},
  \bibnamefont{et~al.}, \bibinfo{journal}{Phys. Rev. Lett.}
  \textbf{\bibinfo{volume}{99}}, \bibinfo{pages}{147207}
  (\bibinfo{year}{2007}), \eprint{arXiv:cond-mat/0702357}.

\bibitem[{\citenamefont{Jungwirth et~al.}(2008)\citenamefont{Jungwirth,
  Gallagher, and Wunderlich}}]{Jungwirth:2008_a}
\bibinfo{author}{\bibfnamefont{T.}~\bibnamefont{Jungwirth}},
  \bibinfo{author}{\bibfnamefont{B.~L.} \bibnamefont{Gallagher}},
  \bibnamefont{and}
  \bibinfo{author}{\bibfnamefont{J.}~\bibnamefont{Wunderlich}}, in
  \emph{\bibinfo{booktitle}{Spintronics edited by
  Tomasz Dietl, David D. Awschalom, Maria Kaminska,  and Hideo Ohno}}
  (\bibinfo{publisher}{Elsevier}, \bibinfo{year}{2008}),
  vol.~\bibinfo{volume}{82} of \emph{\bibinfo{series}{Semicond. Semimet.}}, p.
  \bibinfo{pages}{135}.

\bibitem[{\citenamefont{Jungwirth
  et~al.}(2002{\natexlab{b}})\citenamefont{Jungwirth, Abolfath, Sinova,
  {Ku\v{c}era}, and MacDonald}}]{Jungwirth:2002_c}
\bibinfo{author}{\bibfnamefont{T.}~\bibnamefont{Jungwirth}},
  \bibinfo{author}{\bibfnamefont{M.}~\bibnamefont{Abolfath}},
  \bibinfo{author}{\bibfnamefont{J.}~\bibnamefont{Sinova}},
  \bibinfo{author}{\bibfnamefont{J.}~\bibnamefont{{Ku\v{c}era}}},
  \bibnamefont{and} \bibinfo{author}{\bibfnamefont{A.~H.}
  \bibnamefont{MacDonald}}, \bibinfo{journal}{Appl. Phys. Lett.}
  \textbf{\bibinfo{volume}{81}}, \bibinfo{pages}{4029}
  (\bibinfo{year}{2002}{\natexlab{b}}), \eprint{arXiv:cond-mat/0206416}.

\bibitem[{\citenamefont{{V\'{y}born\'{y}}
  et~al.}(2009)\citenamefont{{V\'{y}born\'{y}}, Kovalev, Sinova, and
  Jungwirth}}]{Vyborny:2008_a}
\bibinfo{author}{\bibfnamefont{K.}~\bibnamefont{{V\'{y}born\'{y}}}},
  \bibinfo{author}{\bibfnamefont{A.~A.} \bibnamefont{Kovalev}},
  \bibinfo{author}{\bibfnamefont{J.}~\bibnamefont{Sinova}}, \bibnamefont{and}
  \bibinfo{author}{\bibfnamefont{T.}~\bibnamefont{Jungwirth}},
  \bibinfo{journal}{Phys. Rev.} \textbf{\bibinfo{volume}{B 79}},
  \bibinfo{pages}{045427} (\bibinfo{year}{2009}), \eprint{arXiv:0810.5693}.

\bibitem[{\citenamefont{Schliemann}(2003)}]{Schliemann:2003_a}
\bibinfo{author}{\bibfnamefont{J.}~\bibnamefont{Schliemann}},
\bibinfo{author}{\bibfnamefont{J. Carlos}~\bibnamefont{Egues}}, \bibnamefont{and}
\bibinfo{author}{\bibfnamefont{D.}~\bibnamefont{Loss}},
  \bibinfo{journal}{Phys. Rev. Lett.} \textbf{\bibinfo{volume}{90}},
  \bibinfo{pages}{146801} (\bibinfo{year}{2003}).

\bibitem[{\citenamefont{Schliemann and Loss}(2003)}]{Schliemann:2003_b}
\bibinfo{author}{\bibfnamefont{J.}~\bibnamefont{Schliemann}} \bibnamefont{and}
  \bibinfo{author}{\bibfnamefont{D.}~\bibnamefont{Loss}},
  \bibinfo{journal}{Phys. Rev.} \textbf{\bibinfo{volume}{B 68}},
  \bibinfo{pages}{165311} (\bibinfo{year}{2003}).

\bibitem[{\citenamefont{Bernevig et~al.}(2006)\citenamefont{Bernevig,
  Orenstein, and Zhang}}]{Bernevig:2006_a}
\bibinfo{author}{\bibfnamefont{B.~A.} \bibnamefont{Bernevig}},
  \bibinfo{author}{\bibfnamefont{J.}~\bibnamefont{Orenstein}},
  \bibnamefont{and} \bibinfo{author}{\bibfnamefont{S.-C.} \bibnamefont{Zhang}},
  \bibinfo{journal}{Phys. Rev. Lett.} \textbf{\bibinfo{volume}{97}},
  \bibinfo{pages}{236601} (\bibinfo{year}{2006}),
  \eprint{arXiv:cond-mat/0606196}.

\bibitem[{\citenamefont{Vurgaftman et~al.}(2001)\citenamefont{Vurgaftman,
  Meyer, and Ram-Mohan}}]{Vurgaftman:2001_a}
\bibinfo{author}{\bibfnamefont{I.}~\bibnamefont{Vurgaftman}},
  \bibinfo{author}{\bibfnamefont{J.~R.} \bibnamefont{Meyer}}, \bibnamefont{and}
  \bibinfo{author}{\bibfnamefont{L.~R.} \bibnamefont{Ram-Mohan}},
  \bibinfo{journal}{J. Appl. Phys} \textbf{\bibinfo{volume}{89}},
  \bibinfo{pages}{5815} (\bibinfo{year}{2001}).

\bibitem[{\citenamefont{Rushforth et~al.}(2009)\citenamefont{Rushforth,
  {V\'{y}born\'{y}}, King, Edmonds, Campion, Foxon, Wunderlich, Irvine,
  {Nov\'{a}k}, {Olejn\'{i}k} et~al.}}]{Rushforth:2007_b}
\bibinfo{author}{\bibfnamefont{A.~W.} \bibnamefont{Rushforth}},
  \bibinfo{author}{\bibfnamefont{K.}~\bibnamefont{{V\'{y}born\'{y}}}},
  \bibinfo{author}{\bibfnamefont{C.~S.} \bibnamefont{King}},
  \bibinfo{author}{\bibfnamefont{K.~W.} \bibnamefont{Edmonds}},
  \bibinfo{author}{\bibfnamefont{R.~P.} \bibnamefont{Campion}},
  \bibinfo{author}{\bibfnamefont{C.~T.} \bibnamefont{Foxon}},
  \bibinfo{author}{\bibfnamefont{J.}~\bibnamefont{Wunderlich}},
  \bibinfo{author}{\bibfnamefont{A.~C.} \bibnamefont{Irvine}},
  \bibinfo{author}{\bibfnamefont{V.}~\bibnamefont{{Nov\'{a}k}}},
  \bibinfo{author}{\bibfnamefont{K.}~\bibnamefont{{Olejn\'{i}k}}},
  \bibnamefont{et~al.}, \bibinfo{journal}{J. Mag. Magn. Mater.}
  \textbf{\bibinfo{volume}{321}}, \bibinfo{pages}{1001} (\bibinfo{year}{2009}),
  \eprint{arXiv:0712.2581}.

\bibitem[{\citenamefont{Vyborny et~al.}(2009)\citenamefont{Vyborny, Kucera,
  Sinova, Rushforth, Gallagher, and Jungwirth}}]{Vyborny:2009_a}
\bibinfo{author}{\bibfnamefont{K.}~\bibnamefont{Vyborny}},
  \bibinfo{author}{\bibfnamefont{J.}~\bibnamefont{Kucera}},
  \bibinfo{author}{\bibfnamefont{J.}~\bibnamefont{Sinova}},
  \bibinfo{author}{\bibfnamefont{A.~W.} \bibnamefont{Rushforth}},
  \bibinfo{author}{\bibfnamefont{B.~L.} \bibnamefont{Gallagher}},
  \bibnamefont{and} \bibinfo{author}{\bibfnamefont{T.}~\bibnamefont{Jungwirth}}
  (\bibinfo{year}{2009}), \eprint{arXiv:0906.3151}.

\bibitem[{\citenamefont{Luo et~al.}(1990)\citenamefont{Luo, Munekata, Fang, and
  Stiles}}]{Luo:1990_a}
\bibinfo{author}{\bibfnamefont{J.}~\bibnamefont{Luo}},
  \bibinfo{author}{\bibfnamefont{H.}~\bibnamefont{Munekata}},
  \bibinfo{author}{\bibfnamefont{F.~F.} \bibnamefont{Fang}}, \bibnamefont{and}
  \bibinfo{author}{\bibfnamefont{P.~J.} \bibnamefont{Stiles}},
  \bibinfo{journal}{Phys. Rev.} \textbf{\bibinfo{volume}{B 41}},
  \bibinfo{pages}{7685} (\bibinfo{year}{1990}).

\bibitem[{\citenamefont{Vasilyev et~al.}(2003)\citenamefont{Vasilyev,
  Suchalkin, Ivanov, Meltser, and Kop'ev}}]{Vasilyev:2003_a}
\bibinfo{author}{\bibfnamefont{Y.~B.} \bibnamefont{Vasilyev}},
  \bibinfo{author}{\bibfnamefont{S.~D.} \bibnamefont{Suchalkin}},
  \bibinfo{author}{\bibfnamefont{S.~V.} \bibnamefont{Ivanov}},
  \bibinfo{author}{\bibfnamefont{B.~Y.} \bibnamefont{Meltser}},
  \bibnamefont{and} \bibinfo{author}{\bibfnamefont{P.~S.}
  \bibnamefont{Kop'ev}}, \bibinfo{journal}{phys. stat. sol.}
  \textbf{\bibinfo{volume}{(b) 240}}, \bibinfo{pages}{R8}
  (\bibinfo{year}{2003}).

\bibitem[{\citenamefont{Ganichev et~al.}(2004)\citenamefont{Ganichev, Bel�kov,
  Golub, Ivchenko, Schneider, Giglberger, Eroms, Boeck, Borghs, Wegscheider
  et~al.}}]{Ganichev:2004_a}
\bibinfo{author}{\bibfnamefont{S.~D.} \bibnamefont{Ganichev}},
  \bibinfo{author}{\bibfnamefont{V.~V.} \bibnamefont{Bel�kov}},
  \bibinfo{author}{\bibfnamefont{L.~E.} \bibnamefont{Golub}},
  \bibinfo{author}{\bibfnamefont{E.~L.} \bibnamefont{Ivchenko}},
  \bibinfo{author}{\bibfnamefont{P.}~\bibnamefont{Schneider}},
  \bibinfo{author}{\bibfnamefont{S.}~\bibnamefont{Giglberger}},
  \bibinfo{author}{\bibfnamefont{J.}~\bibnamefont{Eroms}},
  \bibinfo{author}{\bibfnamefont{J.~D.} \bibnamefont{Boeck}},
  \bibinfo{author}{\bibfnamefont{G.}~\bibnamefont{Borghs}},
  \bibinfo{author}{\bibfnamefont{W.}~\bibnamefont{Wegscheider}},
  \bibnamefont{et~al.}, \bibinfo{journal}{Phys. Rev. Lett.}
  \textbf{\bibinfo{volume}{92}}, \bibinfo{pages}{256601}
  (\bibinfo{year}{2004}).

\bibitem[{\citenamefont{Giglberger et~al.}(2007)\citenamefont{Giglberger,
  Golub, Bel'kov, Danilov, Schuh, Gerl, Rohlfing, Stahl, Wegscheider, Weiss
  et~al.}}]{Giglberger:2007_a}
\bibinfo{author}{\bibfnamefont{S.}~\bibnamefont{Giglberger}},
  \bibinfo{author}{\bibfnamefont{L.~E.} \bibnamefont{Golub}},
  \bibinfo{author}{\bibfnamefont{V.~V.} \bibnamefont{Bel'kov}},
  \bibinfo{author}{\bibfnamefont{S.~N.} \bibnamefont{Danilov}},
  \bibinfo{author}{\bibfnamefont{D.}~\bibnamefont{Schuh}},
  \bibinfo{author}{\bibfnamefont{C.}~\bibnamefont{Gerl}},
  \bibinfo{author}{\bibfnamefont{F.}~\bibnamefont{Rohlfing}},
  \bibinfo{author}{\bibfnamefont{J.}~\bibnamefont{Stahl}},
  \bibinfo{author}{\bibfnamefont{W.}~\bibnamefont{Wegscheider}},
  \bibinfo{author}{\bibfnamefont{D.}~\bibnamefont{Weiss}},
  \bibnamefont{et~al.}, \bibinfo{journal}{Phys. Rev.}
  \textbf{\bibinfo{volume}{B 75}}, \bibinfo{pages}{035327}
  (\bibinfo{year}{2007}).

%\bibitem[{\citenamefont{Dugdale and Basinski}(1967)}]{Dugdale:1967_a}
%\bibinfo{author}{\bibfnamefont{J.~S.} \bibnamefont{Dugdale}} \bibnamefont{and}
%  \bibinfo{author}{\bibfnamefont{Z.~S.} \bibnamefont{Basinski}},
%  \bibinfo{journal}{Phys. Rev.} \textbf{\bibinfo{volume}{157}},
%  \bibinfo{pages}{552} (\bibinfo{year}{1967}). 

\bibitem{Dugdale:1967_a} Matthiessen's rule can, however, be only applied on a
  qualitative level in anisotropic systems. See for example
  J.S.~Dugdale and Z.S.~Basinski, Phys. Rev. {\bf 157}, 552 (1967).



\bibitem[{\citenamefont{Jungwirth et~al.}(2003)\citenamefont{Jungwirth, Sinova,
  Wang, Edmonds, Campion, Gallagher, Foxon, Niu, and
  MacDonald}}]{Jungwirth:2003_b}
\bibinfo{author}{\bibfnamefont{T.}~\bibnamefont{Jungwirth}},
  \bibinfo{author}{\bibfnamefont{J.}~\bibnamefont{Sinova}},
  \bibinfo{author}{\bibfnamefont{K.~Y.} \bibnamefont{Wang}},
  \bibinfo{author}{\bibfnamefont{K.~W.} \bibnamefont{Edmonds}},
  \bibinfo{author}{\bibfnamefont{R.~P.} \bibnamefont{Campion}},
  \bibinfo{author}{\bibfnamefont{B.~L.} \bibnamefont{Gallagher}},
  \bibinfo{author}{\bibfnamefont{C.~T.} \bibnamefont{Foxon}},
  \bibinfo{author}{\bibfnamefont{Q.}~\bibnamefont{Niu}}, \bibnamefont{and}
  \bibinfo{author}{\bibfnamefont{A.~H.} \bibnamefont{MacDonald}},
  \bibinfo{journal}{Appl. Phys. Lett.} \textbf{\bibinfo{volume}{83}},
  \bibinfo{pages}{320} (\bibinfo{year}{2003}), \eprint{arXiv:cond-mat/0302060}.

\bibitem[{\citenamefont{Abolfath et~al.}(2001)\citenamefont{Abolfath,
  Jungwirth, Brum, and MacDonald}}]{Abolfath:2001_a}
\bibinfo{author}{\bibfnamefont{M.}~\bibnamefont{Abolfath}},
  \bibinfo{author}{\bibfnamefont{T.}~\bibnamefont{Jungwirth}},
  \bibinfo{author}{\bibfnamefont{J.}~\bibnamefont{Brum}}, \bibnamefont{and}
  \bibinfo{author}{\bibfnamefont{A.~H.} \bibnamefont{MacDonald}},
  \bibinfo{journal}{Phys. Rev.} \textbf{\bibinfo{volume}{B 63}},
  \bibinfo{pages}{054418} (\bibinfo{year}{2001}),
  \eprint{arXiv:cond-mat/0006093}.

\bibitem[{\citenamefont{Schrieffer and Wolff}(1966)}]{Schrieffer:1966_a}
\bibinfo{author}{\bibfnamefont{J.~R.} \bibnamefont{Schrieffer}}
  \bibnamefont{and} \bibinfo{author}{\bibfnamefont{P.~A.} \bibnamefont{Wolff}},
  \bibinfo{journal}{Phys. Rev.} \textbf{\bibinfo{volume}{149}},
  \bibinfo{pages}{491} (\bibinfo{year}{1966}).

\bibitem[{\citenamefont{Dietl et~al.}(2001)\citenamefont{Dietl, Ohno, and
  Matsukura}}]{Dietl:2001_b}
\bibinfo{author}{\bibfnamefont{T.}~\bibnamefont{Dietl}},
  \bibinfo{author}{\bibfnamefont{H.}~\bibnamefont{Ohno}}, \bibnamefont{and}
  \bibinfo{author}{\bibfnamefont{F.}~\bibnamefont{Matsukura}},
  \bibinfo{journal}{Phys. Rev.} \textbf{\bibinfo{volume}{B 63}},
  \bibinfo{pages}{195205} (\bibinfo{year}{2001}),
  \eprint{arXiv:cond-mat/0007190}.  To 
complete the
link between this reference and Eq.~(\ref{w}), we note that 
$\beta N_0$ shown in Appendix~C of Dietl, Ohno, and Matsukura equals in our
  notation to $J_{pd}n/x$ in a (III$_{1-x}$,Mn$_x$)V material and the
  scattering potential including the proportionality constant reads
  $V=J_{pd}S_{\mathrm{Mn}} \hat{e}_M\cdot \mathbf{s}$ where
  $S_{\mathrm{Mn}}=\frac{5}{2}$ is the total spin of involved Mn
  $d$-electrons and $\mathbf{s}=(s_x,s_y,s_z)$. The proportionality constant
  drops out in the expression for the AMR but it still must provide for the Mn
  impurities to be the dominant source of scattering. Note that the
  proportionality constant is also closely related to $h$ in
  Eq.~(\ref{eq-01d}). 

\bibitem{streda} See e.g. Eq.~(11) in~P.~St\v reda and L.~Smr\v cka,
  phys. stat. sol. (b), {\bf 70}, 537 (1975) which is a more general result
  (finite temperature, more general form of disorder, magnetic field).

\end{thebibliography}

\end{document}